\documentclass[twocolumn]{aastex63}

\newcommand{\Msun}{{\rm M_{\odot}}}

\newcommand{\kpc}{\, {\rm kpc}}

\usepackage{amsmath}
\usepackage{booktabs}
\usepackage{longtable}
\usepackage{graphics}
\usepackage{amstext} 
\usepackage{array}   
\setcitestyle{authoryear}

\submitjournal{ApJS}
\accepted{May 8, 2023}

\shorttitle{BLANK}
\shortauthors{Bautista et al.}

\graphicspath{{../}{figures/}}

\begin{document}

\title{Ultra-Diffuse Galaxies (UDGs) with Hyper Suprime-Cam I: Revised Catalog of Coma Cluster UDGs \footnote{This research is based on data collected at the Subaru Telescope, which is operated by the National Astronomical Observatory of Japan. We are honored and grateful for the opportunity of observing the Universe from Maunakea, which has the cultural, historical, and natural significance in Hawaii.}}
\shorttitle{Revised Catalog of Coma Cluster UDGs}

\author{Jose Miguel G. Bautista}
\affiliation{Department of Physics and Astronomy, Stony Brook University, Stony Brook, NY 11794-3800, USA}
\author{Jin Koda}
\affiliation{Department of Physics and Astronomy, Stony Brook University, Stony Brook, NY 11794-3800, USA}
\author{Masafumi Yagi}
\affiliation{National Astronomical Observatory of Japan, Mitaka, Tokyo, 181-8588, Japan}
\author{Yutaka Komiyama}
\affiliation{National Astronomical Observatory of Japan, Mitaka, Tokyo, 181-8588, Japan}
\affiliation{Department of Advanced Sciences, Faculty of Science and Engineering, Hosei University, 3-7-2 Kajino-cho, Koganei, Tokyo 184-8584, Japan}
\affiliation{Graduate University for Advanced Studies (SOKENDAI), Mitaka, Tokyo 181-8588, Japan}
\author{Hitomi Yamanoi}
\affiliation{Faculty of Science and Technology, Seikei University, 3-3-1 Kichijoji-Kitamachi, Musashino, Tokyo 180-8633, Japan}

\begin{abstract}

This is the first in a series of papers on the properties of ultra-diffuse galaxies (UDGs) in clusters of galaxies. We present an updated catalog of UDGs in the Coma cluster using \textit{g}- and \textit{r}-band images obtained with Hyper Suprime-Cam (HSC) of the Subaru telescope. We develop a method to find UDGs even in the presence of contaminating objects, such as halos and background galaxies. This study expands upon our previous works that covered about half the area of the Coma cluster. The HSC observations covered the whole Coma cluster up to the virial radius and beyond (an area twice larger than the previous studies) and doubled the numbers of UDGs ($r_{\rm eff, r} \geq 1.5$ kpc) and sub-UDGs ($1.0 \leq r_{\rm eff, r} < 1.5$ kpc) to 774 and 729 respectively. The new UDGs show internal properties consistent with those of the previous studies (e.g., S\'ersic index of approximately 1), and are distributed across the cluster, with a concentration around the cluster center. The whole cluster coverage clearly revealed an excess of their distribution toward the east to south-west direction along the cluster center, where Coma connects to the large scale structure, and where a known substructure exists (the NGC4839 subgroup). The alignment of the UDG distribution along the large scale structure around Coma supports the interpretation that most of them lie at the distance of the Coma cluster and the NGC4839 subgroup.
\end{abstract}


\section{Introduction} 

Ultra-diffuse galaxies (UDGs) are extremely large low surface brightness (LSB) galaxies, with effective radii comparable to the Milky Way's, but with  stellar masses as low as $\sim 10^{-2}$ to $10^{-3}$ the Milky Way's (comparable to dwarf galaxies). While LSB galaxies \citep[e.g.,][]{Impey1988, SandageBingelli1984, Dalcanton1997} have been studied for a long time, only a few of them were known to be extremely large LSB galaxies (i.e. UDGs; see \citealt{Yagi2016} Appendix B for a comprehensive literature search for UDGs). Their large abundance has been revealed only recently, which has underlined their importance as a population. Some dwarf galaxy studies found relatively large dwarfs, but not many are as large as UDGs \citep[e.g.,][]{ThompsonGregory1993, Jerjen2000, Conselice2002, Conselice2003, Mieske2007, deRijcke2009, Penny2009, Penny2011}. Previous catalogs occasionally listed the relatively bright centers of UDGs, but identifying their full extents by detecting their diffuse, extended envelopes is often very difficult.

Since the discovery of the 47 in the Coma cluster in \citet{PVD2015}, UDGs have been found to be prolific within many clusters. Coma alone was found to have $\sim 10^3$  UDGs in \textit{R}-band imaging \citep{Koda2015, Yagi2016, Zaritsky2019}. Similar studies have since reported UDGs in other clusters \citep{VDB2016}, including the Virgo \citep{Lim2020}, Hydra \citep{Iodice2020}, Perseus \citep{Gannon2022}, and Fornax clusters, \citep{Venhola2017} as well as in the field \citep{Leisman2017, Jones2018, Greco2018, Zaritsky2019, Barbosa2020, Marleau2021}.

Despite their ubiquity, the nature of such extremely large LSB galaxies remains elusive, and several UDG formation scenarios have been proposed. \citet{PVD2015} initially suggested that UDGs may be ``failed" galaxies, residing in dark matter halos comparable to the Milky Way ($\sim 10^{12}\Msun$), but with truncated star-formation histories. Followup measurements of a representative UDG, DF44, have placed the mass somewhat lower at $\sim 10^{11}\Msun$ \citep{PVD2019, Saifollahi2021}. Another possibility is that UDGs are dwarf galaxies, which have relatively inflated stellar distribution to more conventional dwarfs. \citet{AmoriscoLoeb2016} suggested that UDGs may form as intrinsically diffuse structures in dwarf-mass halos with large angular momenta. Alternatively, the stellar distribution of a dwarf-like progenitor may have been inflated through either internal processes such as stellar feedback \citep{DiCintio2017, Chan2018}, or external processes such as tidal interactions \citep{Carleton2019, Jiang2019, Sales2020, Jones2021} and mergers \citep{Conselice2018, Bennet2018, Wright2021}.

Their environment can be an important factor. Compared to cluster UDGs, UDGs in the field tend to be bluer, more HI-rich, and more irregularly shaped \citep{Greco2018B, Rong2020, Barbosa2020, Kadowaki2021, Tanoglidis2021}. The denser cluster environment may also play a key role in producing quiescent UDGs. Cluster UDGs may initially form as diffuse galaxies in the same manner as the bluer field UDGs, but have undergone additional quenching by the cluster environment \citep{Penny2011, Jiang2019, Prole2019, Grishin2021, Junais2022}. Even within a single galaxy cluster itself, the immediate environment, such as galaxy density, varies and may alter the properties of UDG populations as a function of cluster radii, substructure, and local inhomogeneities. \citet{Janssens2019}, for instance, found that UDGs are asymmetrically distributed around clusters, being deficient in regions of high mass density as traced by gravitational lensing. Furthermore, the UDG distributions appear anticorrelated with the distributions of ultra-compact dwarf (UCD) galaxies, leading them to hypothesize that UCDs are the remnants of destroyed UDGs with surviving compact centers. Thus, the distribution of UDGs relative to the surrounding overdensities and the large scale structure can yield information on how they form and evolve, as different formation mechanisms will leave different imprints on the population distributions and structural parameters.

A complete population study of UDG in a local cluster requires satisfying 3 key points: a large field of view, sufficient point source/surface brightness sensitivities, and high spatial resolution to resolve point sources on top of the UDGs. A large field of view is necessary to cover a wide environmental range, e.g., both the entire cluster and its surroundings. By definition, UDGs are very low surface brightness objects, so a high sensitivity in surface brightness  is needed to both detect and accurately analyze them. An additional complication comes from source confusion: over the large extents of UDGs, there are often foreground and background objects, i.e., stars and background galaxies, as well as their own globular clusters. These objects are compact and typically show higher surface brightnesses (or fluxes within the point spread function) than the UDGs, which hinder the detection and analysis of the UDG. Thus, a high spatial resolution is needed to resolve the sources of confusion and remove them from the UDG.

Identification of LSB objects is inherently difficult, due to the deep depths required to detect them, as well as contamination from other, higher surface brightness objects with LSB tails. In response to these challenges, a number of techniques have been developed to overcome them. One approach is ``thresholding", where clusters of pixels above some threshold level can be associated and identified. For instance, \citet{Bennet2017} spatially-rebinned CFHT images at a target size scale and selected all pixels at some significance above the background to search for diffuse dwarf galaxies around M101. The advantages of this method are that it is relatively simple to implement, and it does not require any prior information on the target object other than the target size. A disadvantage of this is the false positive rate due to background galaxies and the halos around foreground stars. Some assumptions on the properties of the contaminants, such as proximity to bright objects, may be used to improve the detection efficiency. In order to remove the outskirts of bright objects, which can mimic UDGs when detecting large LSB objects, \citet{Greco2018} detected objects at both high and low significance levels. A low significance detection (LSB objects) that shares a certain amount of their area with a high significance detection (high surface brightness object) is associated to the high significance detection as its faint outskirt, and rejected as a UDG. In wide low-resolution observations, where blending is a significant source of contamination, independent high-resolution imaging can be used to model all emission from blended objects in order to be subtracted \citep{PVD2020b}. Improving the signal-to-noise ratio can also be done to improve detection efficiency. \citet{Zaritsky2019}, searching for UDGs in DESI images, used the technique of wavelet transformations to spatially filter the images. Wavelet transformations are similar to smoothing in that they improve the S/N of objects at the size of the kernel, while suppressing the contribution from features at different scales. This allows for UDGs at different kernel sizes to be selectively detected at higher contrast. These technical developments for detection of LSB galaxies are becoming even more important, in view of the upcoming Legacy Survey of Space and Time (LSST) at the Vera Rubin observatory \citep{Ivezic2019}.

In this paper, we revisit UDGs in the Coma cluster. We use new deep \textit{g}- and \textit{r}-band images obtained with Hyper Suprime-Cam (HSC hereafter), and cover the whole Coma cluster out to its virial radius (an area twice larger than previous studies). This catalog is built on \textit{r}-band based detection, as we expect \textit{r}-band to trace the mass distribution better than \textit{g}-band. In order to mitigate the issue of confusion, we develop a procedure to remove contaminants in multiple stages. One of the original selection criteria of UDGs, i.e., the effective radius of $r_{\rm eff, r} \geq 1.5$ kpc, was from an instrumental limitation \citep{PVD2015}, not from an intrinsic property of UDGs. Hence, we catalog all objects with $r_{\rm eff, r} \geq 1.0\kpc$ as UDGs, and call smaller ones ($r_{\rm eff, r} \leq 1.5$ kpc) as ``sub-UDGs". With our larger search area in 2 bands, we roughly double the numbers of UDGs and add color information for all detected objects. With its high resolution and depth, and by covering the entire cluster out to the virial radius and beyond with HSC, our data is  suited for obtaining a more spatially-complete UDG population in the cluster in order to analyze their radial and azimuthal distribution. 

The paper is organized as follows. In Section \ref{sec:term}, we outline our terminology and definiton of UDGs. In Section \ref{sec:HSCimaging}, we discuss our observations of the Coma cluster with HSC. In Section \ref{sec:UDGidentification}, we describe our automated procedure for identifying UDGs, removing background contaminants (small objects) before running a UDG detection algorithm. In Section \ref{sec:Catalog}, we present this catalog and some of their statistical properties. We conclude with a summary in Section \ref{sec:Conclusion}. 

Throughout this paper, we use the following notations:
the apparent magnitude $m$ [mag],
effective radius $r_{\rm eff}$ [kpc],
central and mean surface brightness within $r_{\rm eff}$,
$\mu_{\rm 0}$ and $\langle \mu \rangle_{\rm eff}$ [mag$\cdot$arcsec$^{-2}$],
S\'ersic index $n$,
axis ratio $q$, and position angle $PA$ [degree].
When a band needs to be specified, we will use a subscript,
e.g., $r_{\rm eff, g}$ or $r_{\rm eff, r}$ for $g$- and $r$-band, respectively.
We omit the units for simplicity unless otherwise specified.
The AB magnitude system is used throughout this paper. We adopted cosmological parameters of ($h_0$, $\Omega_M$, $\Omega_{\lambda}$) = (0.697, 0.282, 0.718) \citep{Hinshaw2013} and a distance modulus for the Coma cluster of (m $-$ M)$_{\rm 0} = 35.05$ \citep{Kavelaars2000}. These parameters correspond to a luminosity distance of 102 Mpc, and 0.473 kpc$\cdot$arcsec$^{-1}$ with an angular diameter distance of 98 Mpc. 

\section{Definitions of UDGs and Terminology}
\label{sec:term}

UDGs are large LSB galaxies near the limitation of detection threshold. As such, their exact definition inevitably depends on the limitation of instruments and measurements. An often-used definition is $r_{\rm eff, g} \geq$ 1.5 and $\mu_{\rm 0, g} \geq$ 24.
This $\mu_{\rm 0}$ is not directly from data,
but is from the S\'ersic model assuming an index of $n=1$. 
The equivalent mean surface brightness is $\langle \mu_{\rm g} \rangle_{\rm eff} \geq$ 25.1. 
These are not initially intended as the selection criteria, but are the parameter ranges that enclose the first set of 47 UDGs with the CFHT photometry \citep{PVD2015}.

In the literature, different studies use different criteria due to the instrumental limitations \citep{Yagi2016, VDB2017, Greco2018, Janssens2019, Gannon2022} or astrophysical motivations \citep[i.e. physical properties of the objects; ][]{Zaritsky2019, Lim2020}. Additionally, there are large uncertainties in parameter determination due to existing software and/or to a choice of approach in data reduction and analysis (see Section \ref{sec:UDGCatalog}). Objects that satisfy a set of criteria in one measurement may not satisfy the same criteria in the other measurement (see Section \ref{sec:UDGCatalog}). 
For these reasons, our previous study with Subaru adjusted the thresholds leniently, e.g., $r_{\rm eff}$ from \textsc{SExtractor} \citep{BertinArnouts1996} \citep{Peng2002} to be $>0.7$, instead of $>1.5$, to include all the Dragonfly UDGs in the Subaru measurements \citep{Koda2015, Yagi2016}.
Otherwise, our measurements would have rejected some of the Dragonfly UDGs (see more examples in Sections \ref{subsec:yagiCat}-\ref{subsec:zaritskyCat}).
We also note that the Dragonfly UDGs with \textsc{SExtractor}'s $r_{\rm eff}<1.5$ turned out to have $r_{\rm eff}\geq1.5$ with \textsc{GALFIT} \citep{Peng2002, Peng2010} in \citet{Yagi2016}.

This paper updates our previous Subaru study and largely inherits its approach with some adjustments based on the progress made since then.
Our previous study used \textit{R}-band of the Subaru Prime Focus Camera (Suprime-Cam), instead of \textit{g}-band, for detection and identification
\citep{Koda2015, Yagi2016}. In this study, we adapt \textit{r}-band of Hyper Suprime-Cam (HSC), because the majority of UDGs known so far are red, and the \textit{r}-band should reflect their stellar masses better than $g$-band. 
In addition, a \textit{g}-band identification is more susceptible to some bias: e.g., star-forming and non-star-forming UDGs with the same \textit{g}-band luminosity can have very different \textit{r}-band luminosities and stellar masses.

Following the \citet{Yagi2016}'s spirit for an inclusive catalog and to absorb the uncertainty mentioned above,
we select objects with a threshold of 
\begin{enumerate}
    \item $r_{\rm eff, r}> 1.0$
    \item $\langle \mu_{\rm r} \rangle_{\rm eff} \geq 24$ at a redshift of $z=0$,
\end{enumerate}
in HSC $r$-band.
When necessary, we separate the ones with $r_{\rm eff}=$1.0-1.5 as ``sub-UDGs"
-- this is not for making a subclass of UDGs, but only for a convenience in discussions in this paper.
From this inclusive catalog, readers can select their own objects with their own criteria.

\section{Hyper Suprime-Cam imaging of Coma cluster}
\label{sec:HSCimaging}

\begin{figure}[htp] 
    \centering
    \includegraphics[width=8cm]{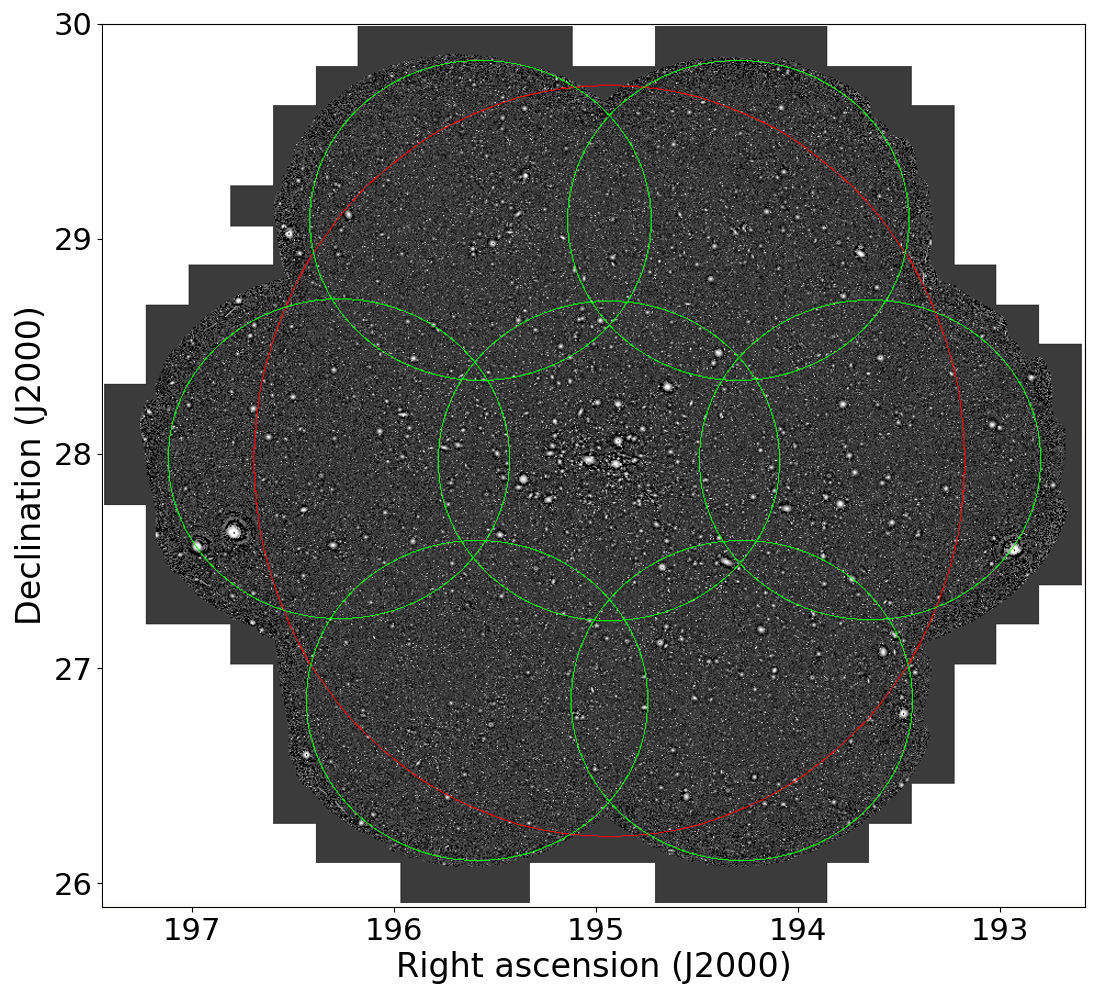}
    \caption{Coverage of the Coma cluster by HSC. The rough footprints of the camera for the 7 pointings are shown in green. The red circle marks 3 Mpc, the virial radius of the cluster, in projected cluster radius.}
    \label{fig:coverage}
\end{figure}

The Coma cluster was observed in \textit{g}- and \textit{r}-bands using Hyper Suprime-Cam (HSC) on the Subaru telescope in March 2016 and in March and June 2017. HSC has 104 CCD detectors, providing a roughly circular field of view with a diameter of $\sim 1.5\arcdeg$ and a scale of $0\arcsec.168$ \citep{Furusawa2018, Kawanamoto2018, Komiyama2018, Miyazaki2018}. The coverage consists of 7 pointings of the camera in a hexagonal pattern (see Figure \ref{fig:coverage}) combined into a single $\approx 15\deg^2$ tract that spans the large cluster up to its virial radius of 3 Mpc ($\approx 1.8\arcdeg$) \citep{Kubo2007}. Typical integration times in \textit{r}- and \textit{g}-bands are about 48 and 136 minutes in total per pointing, respectively, with the dithering pattern optimized so that gaps between CCDs should not overlap. Each dither integration
(called ``visit") was started with the rotator angle of 0 degree, so that the flat pattern near the edge of the field-of-view is consistent among the integrations. The typical seeing throughout the observing runs were $\sim 1\arcsec$. For easier data handling, the tract is divided into square patches of 12' on a side (4200 $\times$ 4200 pixels) in size, with 17" (100 pixels) of overlap between adjacent patches. We search for UDGs on a patch-by-patch basis, and we search 294 patches in total. 

The galactic extinction in \textit{r}-band across the coverage varies between $\sim 0.015$ and $\sim 0.03$ mag according to NASA/IPAC Extragalactic Database (NED)\footnote{https://ned.ipac.caltech.edu/}, which is based on \citet{Schlegel1998} and \citet{SchaflyFinkbeiner2011}. We assume that the variation of the extinction is negligible within each patch, and use the value at the center for all UDGs contained in the patch. Hereafter, the magnitudes and the SB are corrected for the galactic extinction. 

The data were reduced using the HSC pipeline \textit{hscPipe} version 4.0.1 \citep{Bosch2018}, which is built on a software in development for the Legacy Survey of Space and Time (LSST) project \citep{Ivezic2008, Juric2017}, and an additional package for sky-subtraction provided by the HSC Helpdesk. Astrometric and photometric calibrations were done using the Sloan Digital Sky Survey (SDSS)-III DR9 catalog \citep{Ahn2012} distributed with the HSC pipeline, which is re-calibrated by the HSC software team for photometric zero points against those of the Panoramic Survey Telescope \& Rapid Response System 1 \citep{Magnier2013, Schafly2012, Tonry2012}. The background sky was subtracted with a grid of a 512 pixel size ($\sim 87\arcsec$). This angular size is larger than the anticipated size of UDGs ($r_{\rm eff, r}\approx 1.5 $ kpc $\sim 3\arcsec.2$ at the distance of Coma) and the one used in the previous studies $\sim 51\arcsec$ \citep{Koda2015, Yagi2016}. 
We take the fully reduced, photometrically and astrometrically calibrated, sky-subtracted images from each visit, and make median-stacked images using \textit{imcio2} \citep{Yagi2002} to reduce the presence of bright artifacts from single exposures.

The HSC pipeline measures the photometric zero point
with 24-pixel aperture and derives the amount of aperture correction.
The scatters in aperture correction among the visits are only 1.2\% in $r$-band
and 0.8\% in $g$-band.
Hence, we applied constant aperture corrections, -0.026 mag for \textit{r}-band and -0.025 mag for \textit{g}-band, to all patches.

\section{Identification of UDGs}
\label{sec:UDGidentification}

Historically, LSB objects tended to be overlooked because of the inherent difficulty in detecting extended objects only a few percent brighter than the background sky. 
Several faint background objects are often on top of UDGs and are a major source of confusion. We therefore need to develop a UDG identification procedure to resolve these issues. 
We conduct the search on image patches, hereafter ``patches" (Section \ref{sec:HSCimaging}), and extensively use \textsc{SExtractor} version 2.19.5 \citep{BertinArnouts1996} and \textsc{GALFIT} version 3.0.5 \citep{Peng2002, Peng2010}. 
For some of the steps below, we use small cutout images, hereafter ``cutouts", with a size of 53."9 $\times$ 53."9 (321 $\times$ 321 pixels) around the barycenters of the UDGs, as measured by \textsc{SExtractor} after removing contaminants in a separate ``cleaning" stage. 
The broad outline is as follows, with more detailed descriptions presented in the following subsections:

\begin{enumerate}
    \item \textit{Removing bright objects and compact objects}. 
    When compact objects are on top of a UDG, \textsc{SExtractor} tends to split the UDG and assign the pieces to the compact objects as their diffuse tails. Therefore, we first clean the patches by removing the compact objects with \textsc{SExtractor} and the unsharp masking technique (Section \ref{sec:removecompact}). At this stage, only the bright cores of the compact objects are removed, while their diffuse tails remain in the images.
    
    \item \textit{Generating crude candidate catalog}.
    We run \textsc{SExtractor} on the cleaned \textit{r}-band patches and select UDGs based on the \textsc{SExtractor} parameters. Here we generate a blanket candidate catalog that, in the parameter space, includes all objects in the volume larger than the part where UDGs are expected to occupy
    (Section \ref{sec:sourcedetect}).
    This procedure leaves false detections, mostly the blends of and contamination by the tails of the cleaned compact objects. 
    
    \item \textit{Removing blended objects}. 
    We identify false detections due to the blending and contamination of the compact objects on a small cutout of raw (uncleaned) image for each UDG. We again utilize \textsc{SExtractor} to (over)split the UDG into pieces and to see if the pieces are associated with compact objects. We remove the UDGs from the catalog when they are blended or heavily contaminated by the compact objects (Section \ref{sec:remblend}).
    
    \item \textit{Refinement of initial parameters and catalog selection}. We use \textsc{GALFIT} to refine the structural parameters of the UDGs with a 2-dimensional S\'ersic function. In order to avoid background contamination, we use the cleaning algorithm from Step 1 (Section \ref{sec:removecompact}) to generate a mask for compact objects. We then visually inspect each cutout, remove any clear false positive including stellar halos and tidal tails, and perform any necessary adjustments to the mask. After removing false positives, we finalize the catalog selection, selecting all objects that satisfy the definition of a UDG by measured \textsc{GALFIT} parameters (Section \ref{sec:galmodel}).
    
    \item \textit{Final parameter refinement}. Finally, we refine the fits by taking into account the possibility of a nuclear component. To evaluate the possibility, we re-fit the candidates, this time including a PSF function. The magnitude of any possible PSF is unknown, so we search the space of initial guesses in PSF magnitude ($m_{\rm psf}$). We compare the fits with the single S\'ersic and S\'ersic + PSF in terms of the contrast between the S\'ersic and PSF aperture fluxes at the center and Bayesian information criterion, and determine whether the single S\'ersic or S\'ersic + PSF fit result is appropriate (Section \ref{sec:finalRefinement}).
    
\end{enumerate}

\subsection{Removing Bright Objects And Compact Objects} \label{sec:removecompact}

\begin{figure*}[htp] 
    \centering
    \includegraphics[width=18cm]{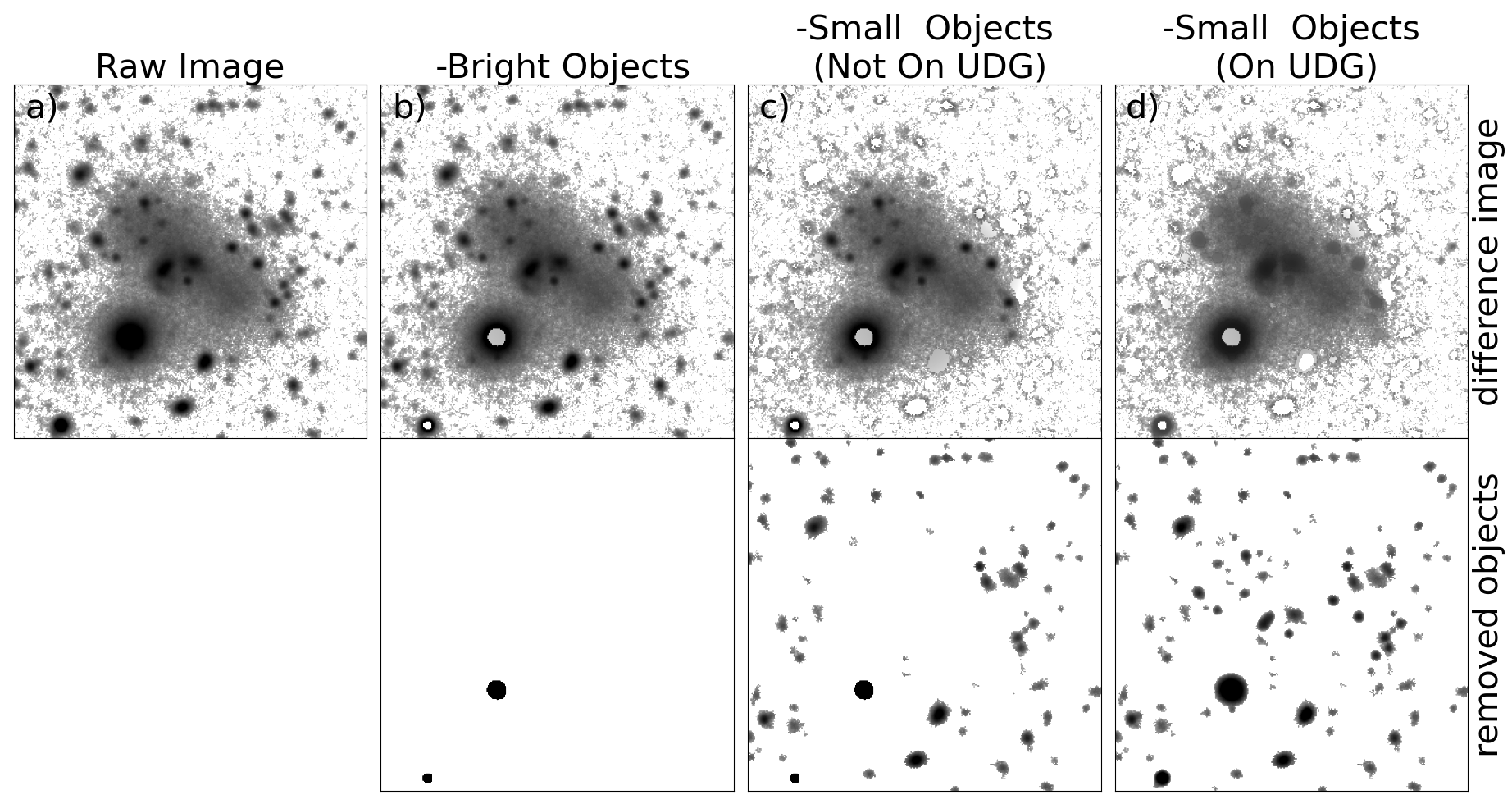}
    \caption{Step-by-step demonstration of our cleaning procedure for one UDG. (top) From left to right: none (original raw image), high brightness objects, smaller objects in isolation, and smaller objects on top of the UDG. (bottom) The removed objects.}
    \label{fig:cleaningLayers}
\end{figure*}

Because of their large extents and low brightnesses, the search for UDGs with \textsc{SExtractor} encounters two major problems: oversplitting and blending. Multiple compact objects in the foreground and background often overlap with UDGs. When \textsc{SExtractor} detects those objects on UDGs, it tends to split the diffuse extended light of the UDGs into multiple pieces and associate them to the foreground/background objects (oversplitting). In addition, UDGs located next to bright stars or galaxies can be confused with the tails of those brighter neighbors (blending). \textsc{SExtractor} is equipped with two control parameters to tackle these problems, \textit{DEBLEND\_MINCONT} and \textit{DEBLEND\_NTHRESH}, but we found it impossible to optimize them to rescue UDGs simultaneously against the splitting and blending. Therefore, we leave the two parameters at default settings and take a different approach. 

Our approach is to remove the contaminating objects before running \textsc{SExtractor} for UDG detections. We aim to detect and remove as many contaminants as possible without disturbing the portions of the images containing UDGs. One method is to use separate \textsc{SExtractor} runs, each optimized to detect a type of contaminant separately \citep{Rix2004, Barden2012, Prescott2012, Greco2018}. We exploit the fact that UDGs are by definition both large and faint, and hence, detect and remove all objects that are either too small or too bright. This cleaning consists of \textsc{SExtractor} runs on the patches, each of which is tailored to find different types of contaminants: objects with brightness too high to be UDGs, and objects too small to be UDGs. We identify these types of objects with \textsc{SExtractor} by adjusting the control parameters (described below). For each type, \textsc{SExtractor} outputs the CHECKIMAGE image that contains only the detected objects. These images are subtracted from the original image. Figure \ref{fig:cleaningLayers} shows the procedure in sequence.

We first identify and remove objects brighter than UDGs from the patches (see Figure \ref{fig:cleaningLayers}b). In \textsc{SExtractor}, the bright objects can be detected by setting a higher detection threshold (23 mag$\cdot$arcsec$^{-2}$) without limits on size. The halos of bright objects will remain in the image and will be removed at catalog level in Section \ref{sec:sourcedetect}. 
The masking fraction due to bright objects in each patch is typically 5\%, but it can be as high as 10\% within $\sim$0.2 degrees of the cluster center.

The second type of contaminant, objects smaller than UDGs, can be split into 2 cases: smaller objects in isolation (i.e., outside the UDGs; Figure \ref{fig:cleaningLayers}c), as well as ones on top of UDGs (Figure \ref{fig:cleaningLayers}d). These two cases have to be treated separately in practice. The smaller objects in isolation can be detected by setting a maximum area threshold (\textit{DETECT\_MAXAREA}) of 400 pixels (11.29 arcsec$^2$) at a detection threshold of 27.5 mag$\cdot$arcsec$^{-2}$ ($\sim 1.5 \sigma$ detection threshold across all patches). Note for comparison, a UDG ends up having a typical area on the order of 2000 pixels (56.4 arcsec$^2$). To find the smaller objects on top of UDGs, we use the unsharp masking technique. In this technique, we smooth the original \textit{r}-band patch by a convolution with a Gaussian kernel (FWHM = 2."6), and subtract this smoothed copy from the original. The size of the kernel was determined by trial and error to prevent oversplitting. What remains are the features with sizes on the order of, or smaller than, the smoothing kernel. Thus, the UDGs are mostly removed, and the overlapping compact objects are still left in the patch. We run \textsc{SExtractor} on this unsharp masked patch with the maximum area threshold (\textit{DETECT\_MAXAREA}) set at 400 pixels (11.29 arcsec$^2$). 
The masking fraction due to compact objects in each patch is typically 8\%. 

Some contaminants still remain in the cleaned image (Figure \ref{fig:cleaningLayers}d). A more stringent threshold can remove many of the remaining contaminants, but it can also damage the UDG-part of the image significantly. In practice, the remaining contaminants are relatively minor and do not hinder the detection and analysis of UDGs.

\begin{figure*}[htp]
    \centering
    \includegraphics[width=8cm]{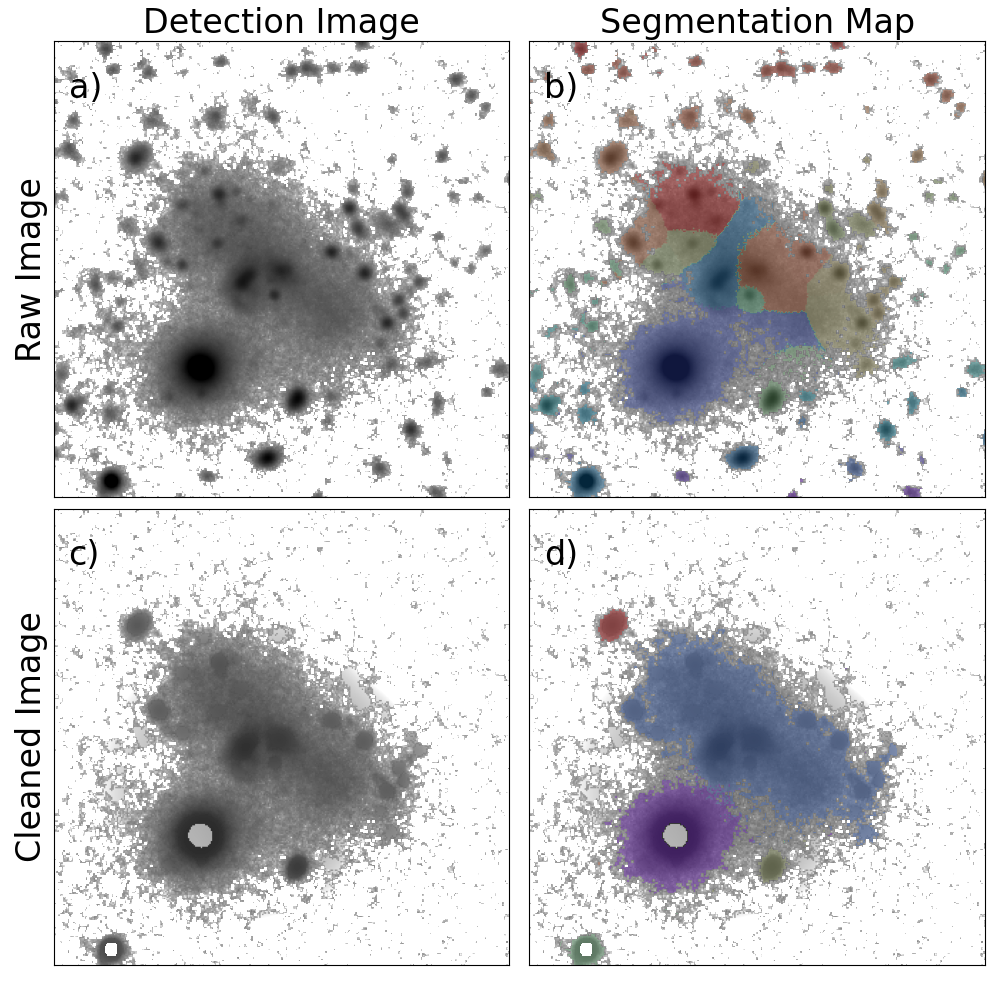}
    \caption{\rm Effect of cleaning on the ability of \textsc{SExtractor} to detect a UDG intact. (a) Raw image of a UDG before cleaning. (b) Corresponding segmentation map overlaid on the raw image. Different objects found by \textsc{SExtractor} are separated by color. The UDG is oversplit by the presence of objects on and around it.(c) Raw image after cleaning. (d) Corresponding segmentation map overlaid on the cleaned image. The UDG is now detected as a single object.}
    \label{fig:cleaningEffect}
\end{figure*}

\begin{figure}[htp]
    \centering
    \includegraphics[width=8cm]{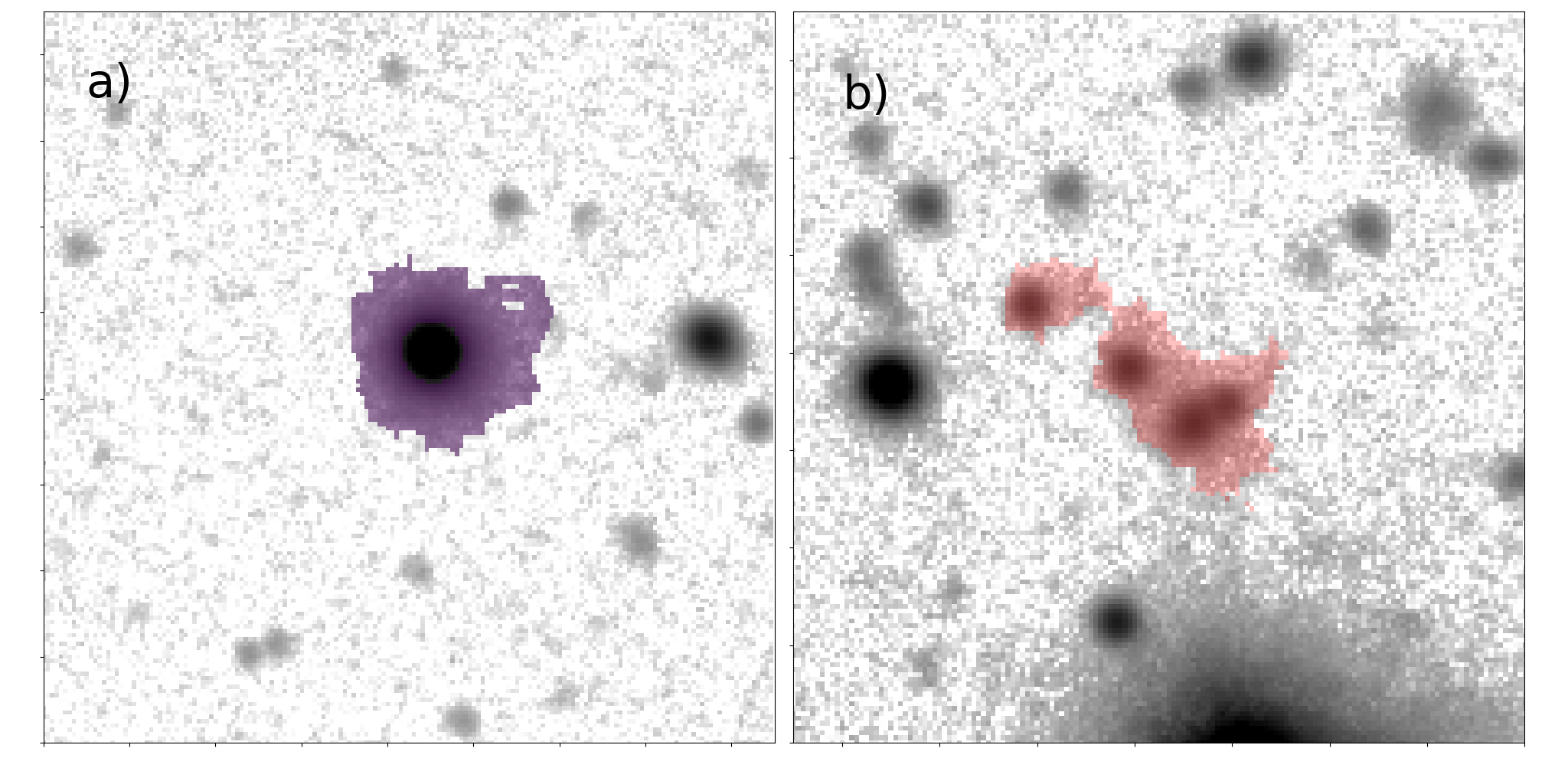}
    \caption{Example contaminants left by cleaning. (a) A bright object whose outer emission is detected as a UDG. (b) An aggregation of compact objects that was detected as a UDG due to blending.}
    \label{fig:autocontaminant}
\end{figure}

These operations remove the peaks of contaminating objects, preventing oversplitting (Figure \ref{fig:cleaningEffect}). However, the faint outskirts of the removed objects still remain in the image. The remnants include halos of bright objects and blended tails of clusters of small objects (Figures \ref{fig:autocontaminant}a and b respectively). \textsc{SExtractor} detects the blended stellar clusters as large and faint objects, which mimic UDGs in the eyes of \textsc{SExtractor}. These must be removed by other means in the following steps.

\subsection{Generating Crude Candidate Catalog} \label{sec:sourcedetect}

The previous step removes most of the contaminating objects, including the ones on top of UDGs, which prevents the oversplitting of the UDGs by \textsc{SExtractor}. However, these cleaned patches still have the halos of the removed bright objects (Figure \ref{fig:cleaningEffect}d (purple)), which still results in false detections by \textsc{SExtractor}. We choose to remove the halos after making an initial catalog of UDGs with \textsc{SExtractor}, instead of attempting to remove them from the image before making the catalog.

For each patch, we run \textsc{SExtractor} in the dual imaging mode to detect UDGs and to measure their parameters: we detect on the cleaned patches, and run analysis on the corresponding uncleaned patches. In the detection image, \textsc{SExtractor} may identify the halo of removed bright objects as a UDG, but the measurement in the analysis image includes the removed bright peaks. Hence, the object will has a high surface brightness and will be rejected via a surface brightness cut below. 

With the \textsc{SExtractor} outputs, we make parameter cuts to filter out non-UDGs. We tune the cutoffs in the parameter space to include as many UDGs from the \citet{Yagi2016} catalog as possible. We select preliminary UDGs by the following parameter cuts in the \textit{r}-band:
\begin{enumerate}
    \item The FLAGS value is less than 32.
    \item The Petrosian radius ($PETRO\_RADIUS$) is measured (non-zero).
    \item The apparent magnitude ($MAG\_AUTO$) is brighter than 26. 
    \item The half light radius ($FLUX\_RADIUS$ with $FLUX\_FRAC=0.5$) is greater than 1.375 arcsec (approximately 0.65 kpc). This is smaller than the definition of a UDG ($r_{\rm eff} \geq 1.5 \kpc$), and will be refined once more accurate parameters for the objects are obtained later in Section \ref{sec:galmodel}.
    \item The full width half maximum ($FWHM $) is greater than 2.65 arcsec (approximately 1.25 kpc).
    \item The isophotal area ($ISOAREA\_IMAGE$) above the analysis threshold (27.5) is greater than 840 pixels (23.71 arcsec$^2$). This corresponds to a circular area of radius 1.3 kpc.
    \item The mean surface brightness within the effective radius ($MU\_MEAN\_MODEL$) is between 23.44 and 28.88. The bright end of this cut is brighter than the cut we will ultimately use for the catalog ($\langle \mu_{\rm r} \rangle_{\rm eff} \geq 24$), and will be refined once more accurate parameters for the objects are obtained later in Section \ref{sec:galmodel}. 
    \item The difference between the surface brightnesses at the effective radius ($MU\_EFF\_MODEL$) and $\langle \mu_{\rm r} \rangle_{\rm eff}$ is less than 1.44. 

\end{enumerate}

Criteria 1 and 2 help remove spurious detections. 

Criterion 3 also removes noises with the faintest magnitude cut ($m_{\rm r} <$  26). 

\begin{figure}[htp] 
    \centering
    \includegraphics[width=8cm]{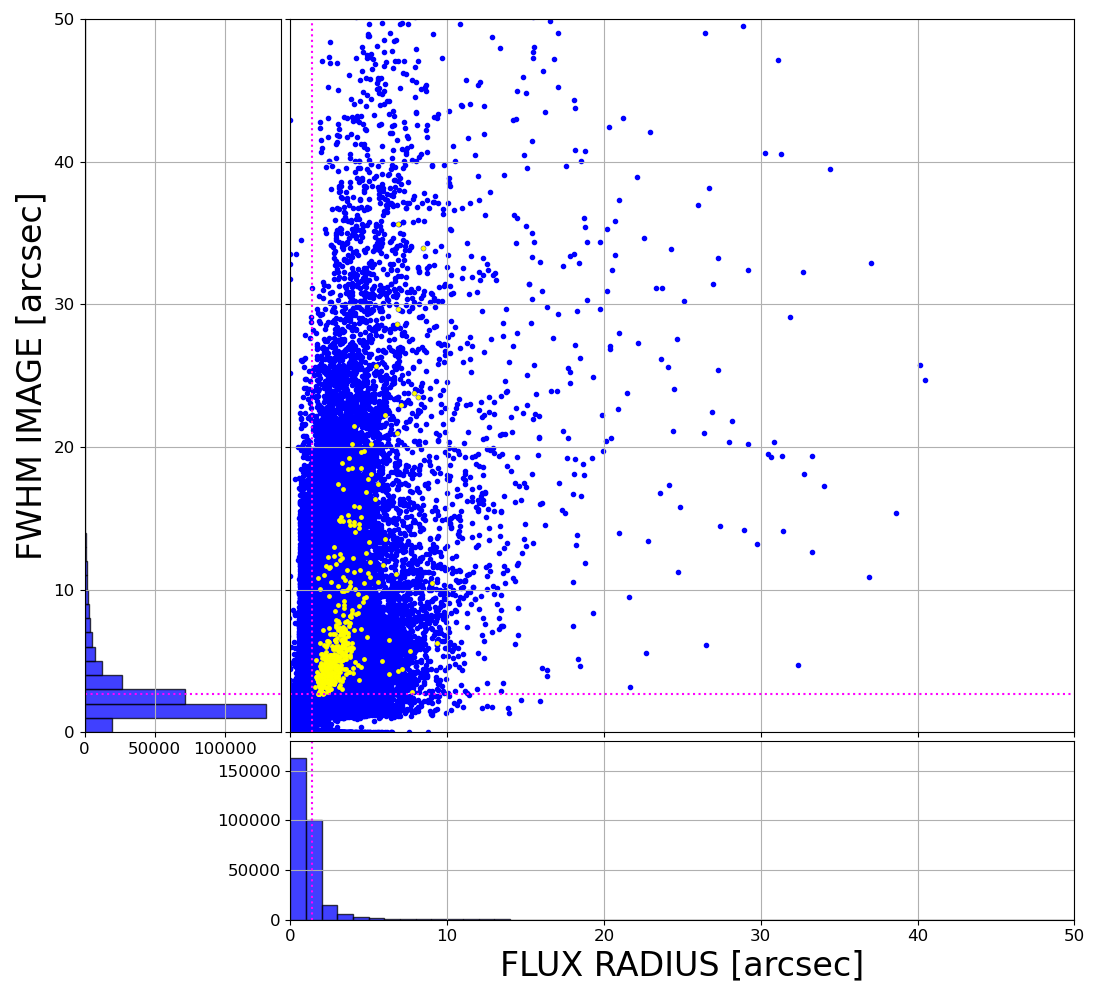}
    \caption{Comparison between \textsc{SExtractor} measured full-width half maximum (\textit{FWHM}) and the half light radius (\textit{FLUX\_RADIUS} with \textit{FLUX\_FRAC} = 0.5) for all detected objects (blue). Overplotted are catalog objects in the limited surface brightness range $25.5 \leq \langle \mu_{\rm r} \rangle_{\rm eff} \leq 26$ (yellow), to demonstrate the distribution of UDGs. Criteria 4 and 5 are the magenta lines.}
    \label{fig:fwhmFluxRad}
\end{figure}

Criteria 4 and 5 may appear somewhat redundant, but we found this combination work for our purpose after trials and error. Figure \ref{fig:fwhmFluxRad} shows all UDGs detected by \textsc{SExtractor} in blue, while yellow and red markers show the objects in our final UDGs catalog. These two criteria clearly eliminates two distinct, abundant populations of non-UDGs. 

Criterion 6 is a measure of the size of the galaxy, independent of the shape of the object or model used to fit it.

Criterion 7 removes spurious detections below the detection limit in surface brightness ($\langle \mu_{\rm r} \rangle_{\rm eff} \geq 28.88$ mag$\cdot$arcsec$^{-2}$), and also removes bright objects ($\langle \mu_{\rm r} \rangle_{\rm eff} \leq$ 23.44). The bright limit is set relatively less stringent as to keep the UDGs catalog inclusive (any marginal objects will be removed later). 

Criterion 8 removes the objects with extremely steep surface brightness profiles. Assuming a S\'ersic profile, this corresponds to a constraint of the S\'ersic index $n \lesssim 4$. 

These cuts leave a catalog of detected candidates of 8950 objects. To reiterate, this is a crude catalog, which intentionally contains many entries that do not qualify as UDGs. 

\subsection{Identifying Blended Objects} 
\label{sec:remblend}

We searched for UDGs in the cleaned patches (Section \ref{sec:removecompact}) and generated the UDG catalog (Section \ref{sec:sourcedetect}). The cleaning process is developed to avoid the oversplitting problem when compact objects are on top of UDGs. Once the compact objects are cleaned, \textsc{SExtractor} can find UDGs as single objects, but it now causes an over-blending problem when multiple small objects are clustered; their halos, after the cleaning (peak subtractions), can appear connected and are identified as a single object (Figure \ref{fig:autocontaminant}b). In this step, we re-analyze the cutouts of individual UDGs and remove the contamination. We again utilize \textsc{SExtractor}.

This time, we aim to distinguish clustered compact objects from UDGs. We investigate cutouts from the \textit{uncleaned} original patches, in which compact objects are not removed. We let \textsc{SExtractor} split a UDG into small pieces in a cutout (as in Figure \ref{fig:cleaningEffect}b). We check the properties of each piece to see if it consists only of compact objects, or if there is an excess of background emission in addition to the compact objects. We identify pieces consisting of only compact objects as those that satisfy two criteria; \textit{CLASS\_STAR} $\geq$ 0.1 and \textit{ISOAREA} smaller than 350 pixels (9.87 arcsec$^2$). The area of pieces consisting of only compact objects is subtracted from the total area of the UDG. If the remaining area is still greater than or equal to 350 pixels, we keep it in the UDG catalog. %This leaves 5686 ($\rightarrow$ 6026) objects. 

Some objects are located between adjacent patches and have 2 (or more) entries. We identify those duplicates based on their coordinates and average their images into a single entry. This leaves 5581 objects.

\subsection{Initial Parameter Refinement And Catalog Selection} \label{sec:galmodel}
We use \textsc{GALFIT} to identify UDGs with refined parameter measurements in $m$, $r_{\rm eff}$, $n$, $q$, and $PA$. After the selection in \textit{r}-band, we run \textsc{GALFIT} to derive the \textit{g}-band parameters.

\subsubsection{Local Sky-subtraction and S\'ersic Modeling}
\label{subsec:localSky}
\begin{figure*}[htp] 
    \centering
    \includegraphics[width=18cm]{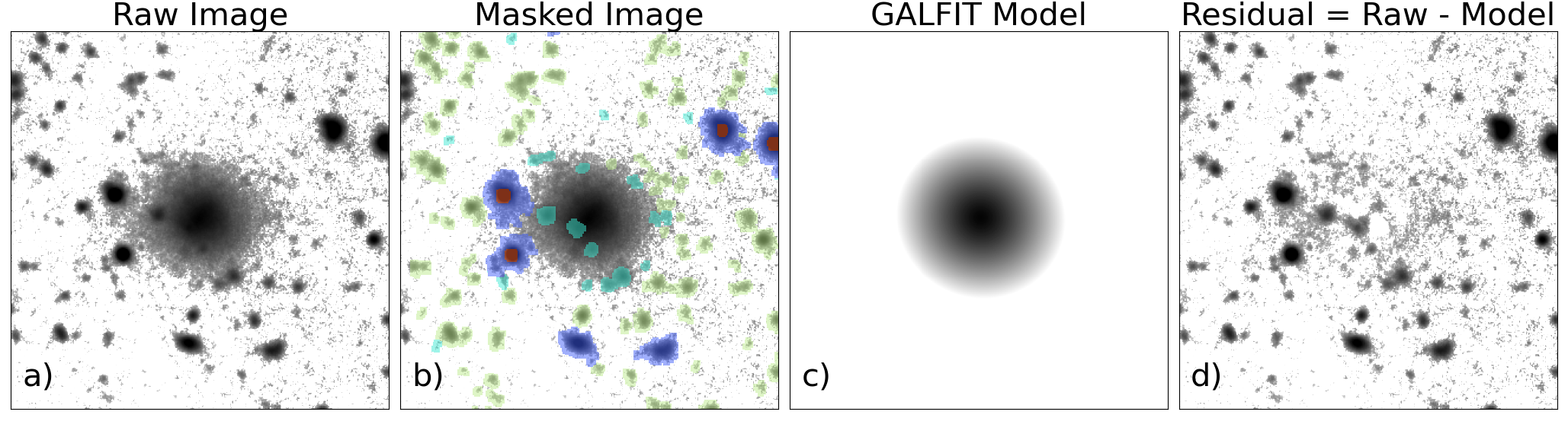}    
    \caption{(a) Raw cutout of a UDG, (b) raw cutout with mask of each contaminant overlaid (color-coded), (c) modeled 2d S\'ersic profile from \textsc{GALFIT}, and (d) residual image after subtracting \textsc{GALFIT} model from raw image.}
    \label{fig:modeling}
\end{figure*}
First, we re-subtract the local sky from a cutout because a small error in the sky determination impacts on the properties of faint UDGs. The sky is estimated as ``2.5$\times$median - 1.5$\times$mean" of the cutout flux \citep{DaCosta1992}. This flux excludes the circular region within a 10$\arcsec$ radius from the center of the UDG to avoid the light of the UDG. The center of the UDG is redefined with the \textsc{SExtractor} measured center after cleaning. To exclude the contaminants, we employ the mask made in Section \ref{sec:removecompact}, but with a modification of expanding the masked regions by 3 pixels outward to include the halos and tails around the masked objects. Figure \ref{fig:modeling}b shows an example of the mask for a sample UDG.

We then fit a single-component 2-d S\'ersic profile to each UDG cutout and extract their structural properties. An initial guess for each \textsc{GALFIT} parameter is made based on the \textsc{SExtractor} outputs. The PSF image is generated from running \textsc{PSFEX} version 3.22.1 \footnote{https://www.astromatic.net/software/psfex/}\citep{Bertin2013} on the patch the cutout is taken from. We use a constant sigma image, assuming the dominant source of noise is sky-based. Figures \ref{fig:modeling}c and d also show the resulting \textsc{GALFIT} model and residual for a sample UDG. The \textit{g}-band S\'ersic index is also held fixed at the \textit{r}-band value. 

\subsubsection{Artifact Removal}
\begin{figure}[htp]
    \centering
    \includegraphics[width=8cm]{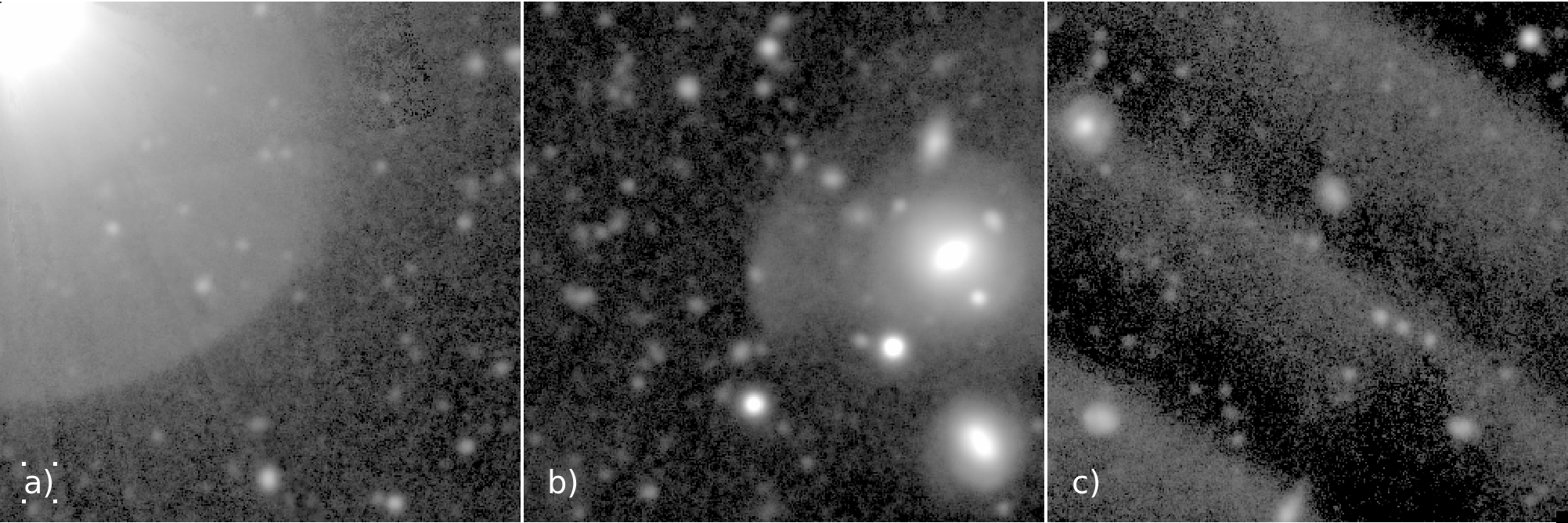}
    \caption{Examples of objects removed during visual inspection. (a) A stellar halo, where the source star is beyond the left edge of the cutout. (b) A tidal feature associated to the large galaxy to the right top of the cutout. (c) Optical ghosts.}
    \label{fig:VIcontaminant}
\end{figure}

This \textsc{GALFIT} run gives solutions for most UDGs, but not for 920 of 5581 cases. Of the 920 objects, 629 turned out to be either artifacts or spurious detections (see Figure \ref{fig:VIcontaminant}) and do not have convergent \textsc{GALFIT} solutions. These 629 were identified and removed manually through visual inspection. In principle, this process can also be automated, but the number of the non-convergent cases is relatively small, and the spurious detections are obvious in the cutouts, and we chose to remove them manually. Each cutout was examined, and removed if its morphology was in the following categories (Figure \ref{fig:VIcontaminant}):

\begin{enumerate}
    \item \textit{Stellar halos}: arc shaped edges with a sharp drop in intensity around very bright stars.
    
    \item \textit{Tidal features}: diffuse and irregular structures, or shell structures, around the vicinity of or connected to brighter galaxies.
    
    \item \textit{Optical Ghosts}: sharp, typically thin lines running across the image. Most of them are due to reflections inside the camera and telescope.
\end{enumerate}

The remaining cases did not converge to a solution, mainly due to an incomplete masking of the contaminants. In such cases, the mask is manually dilated to remove the contaminants where possible. This step leaves 4952 objects.

\subsubsection{\textsc{GALFIT}-based Selection}
\label{sec:galSelection}
After removing the artifacts, we make the final selection cuts (see Section \ref{sec:term}):
\begin{enumerate}
    \item $r_{\rm eff, r} \geq$ 1.0,
    \item $\langle \mu_{\rm r} \rangle_{\rm eff} \geq$ 24.1 at Coma ($\geq 24.0$ at $z=0$),
    \item \textit{color within 2 kpc radius}: (\textit{g}-\textit{r})$_{\rm 2 kpc}$ $\leq$ 1.0.
\end{enumerate}
Note again that our notations are
the effective radius $r_{\rm eff}$ [kpc],
the central and mean surface brightness within $r_{\rm eff}$,
$\mu_{\rm 0}$ and $\langle \mu \rangle_{\rm eff}$ [mag$\cdot$arcsec$^{-2}$],
the S\'ersic index $n$,
axis ratio as $q$, and position angle $PA$ [degree].

\begin{figure}
    \centering
    \includegraphics[width=8cm]{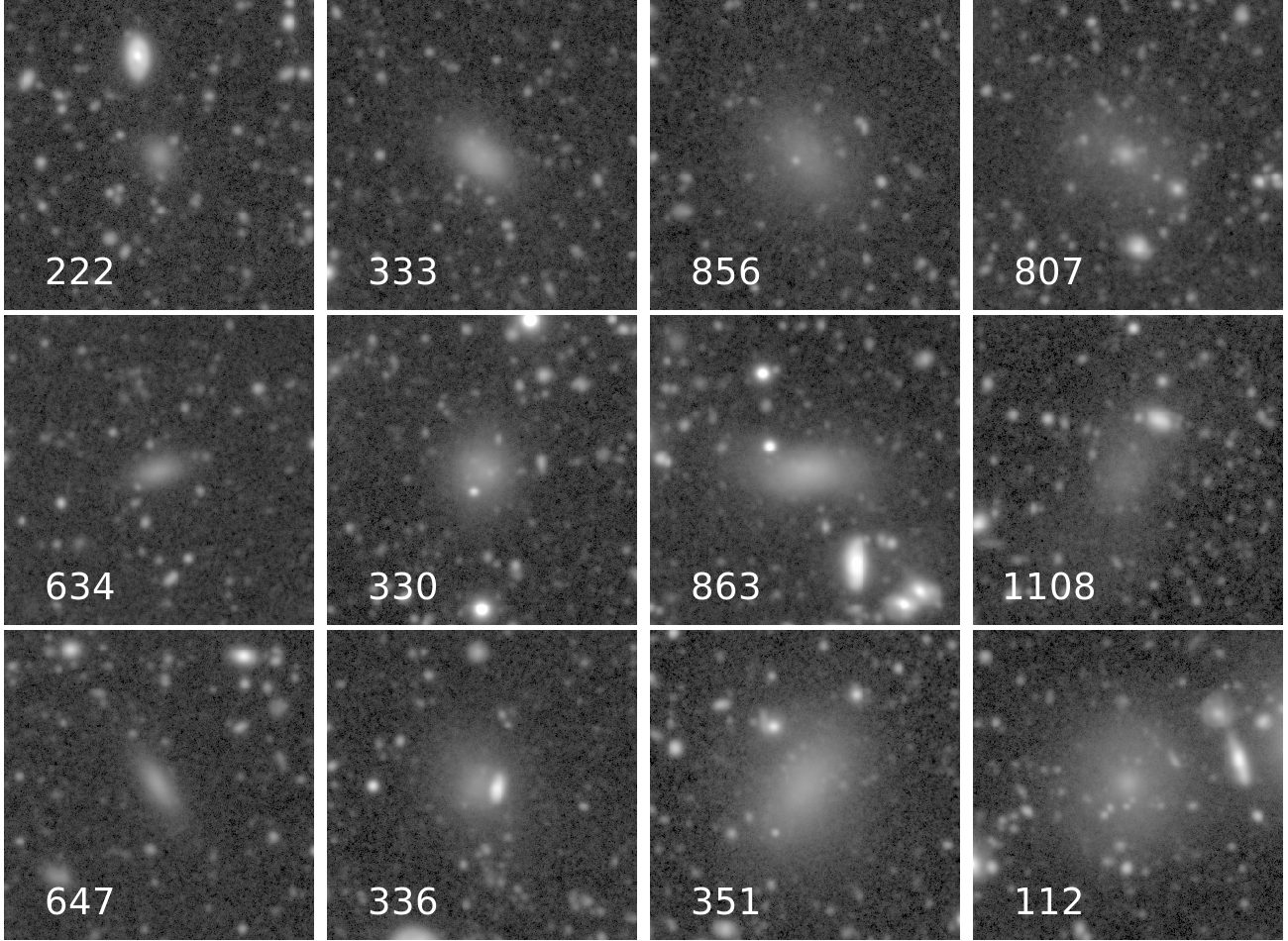}
    \caption{Example UDGs organized by $r_{\rm eff, r}$. Columns from left to right: UDGs larger than 1, 2, 3, and 4 kpc. DF44 is shown in the bottom of the 3rd column (ID = 351). Each UDG's ID number of in this catalog is shown in the bottom right of each cutout.}
    \label{fig:reffSample}
\end{figure}

Criterion 1: 
as mentioned in Section \ref{sec:term}, our UDGs include ``UDGs" with the common threshold of $r_{\rm eff} \geq 1.5$, and ``sub-UDGs" with $r_{\rm eff}=1.0$-$1.5$. The latter are included to mitigate the large error in $r_{\rm eff}$ and to avoid missing true UDGs by this error.
Figure \ref{fig:reffSample} shows sample UDGs organized by $r_{\rm eff, r}$ for comparison. 

Criterion 2:
following \citet{Yagi2016}, we adopt $\langle \mu \rangle_{\rm eff} \geq 24.0$.
We take this as a definition at $z=0$, and the cosmological dimming 
changes the threshold to 24.1 for Coma ($z=0.023$).

Here, we neglect the band difference between HSC $r$ and Suprime-Cam $R$ in \citet{Yagi2016}.
Using the SDSS spectra of galaxies around the Coma cluster within a radius of $2\arcdeg$ and in $0.015<z<0.035$,
we derive the conversion equation \citep[see][]{Yagi2013}:

\begin{equation} \label{eqR}
    \begin{split}
        R_{\rm Suprime} = r_{\rm HSC} - 0.1056(g-r)_{\rm HSC} - 0.01676.
    \end{split}
\end{equation}

For a median color of $(g-r)_{\rm HSC}\sim 0.55$ among the UDGs (Section \ref{sec:color}), $R=r-0.07$.
\citet{Yagi2016} did not take into account the cosmological dimming.
Hence, our inclusion of the dimming correction and neglect of the color conversion more or less compensate,
and the two studies use roughly the same threshold.

Comparisons of this threshold to those of the other studies require some conversions. We adopt the $\langle \mu \rangle_{\rm eff}$ instead of $\mu_{\rm 0}$ because $\langle \mu_{\rm r} \rangle_{\rm eff}$ is less sensitive to the S\'ersic index $n$. Many studies used $\mu_{\rm 0}$ of the S\'ersic model, not of a UDG image, under an assumption of $n=1$, and in this case,

\begin{equation}
    \langle \mu \rangle_{\rm eff} = \mu_{\rm 0} + 1.124.
\end{equation}

\begin{figure}
    \centering
    \includegraphics[width=8cm]{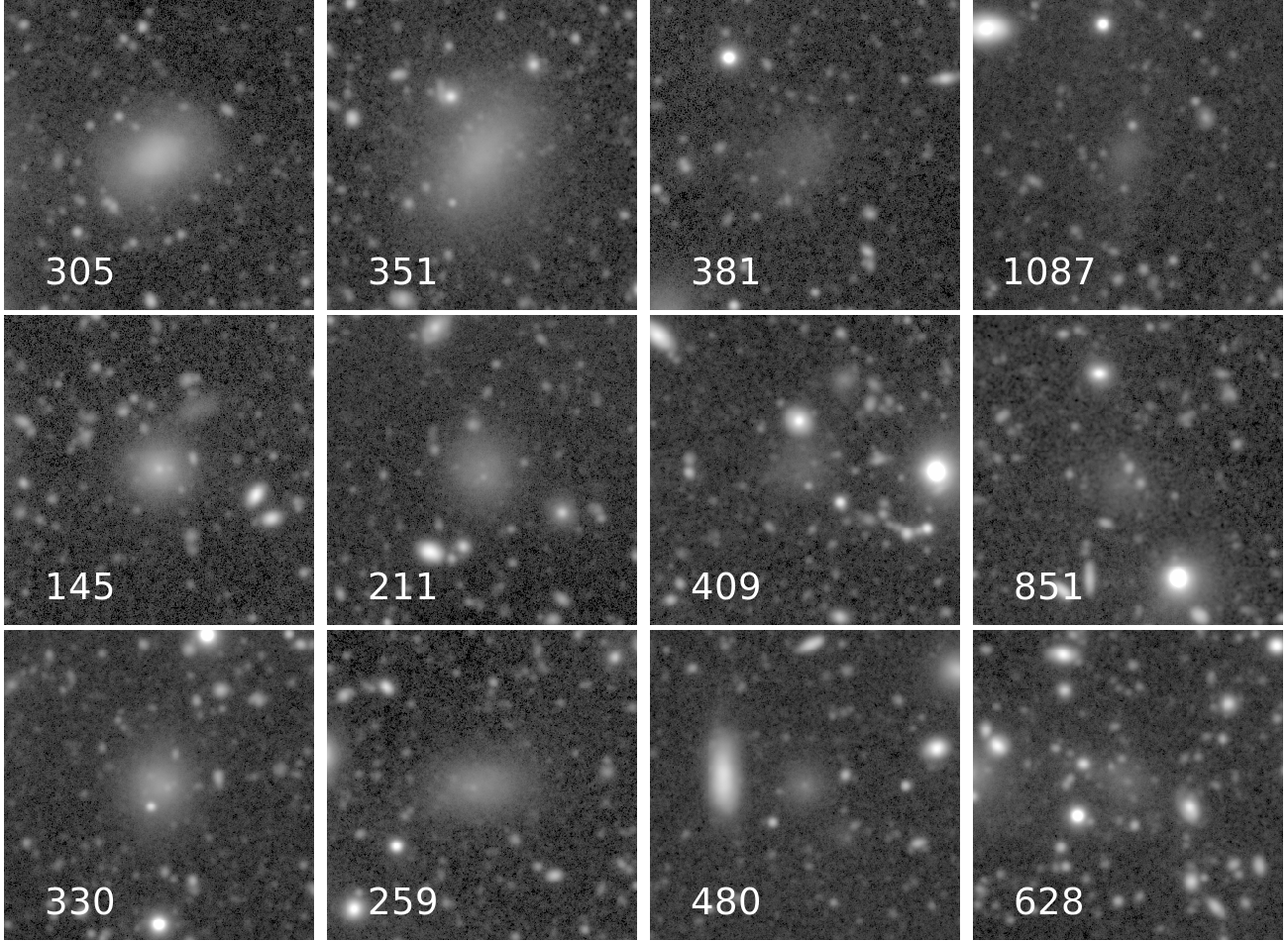}
    \caption{Example UDGs organized by $\langle \mu_{\rm r} \rangle_{\rm eff}$. Columns from left to right: UDGs fainter than 24, 25, 26, and 27 mag$\cdot$arcsec$^{-2}$. DF44 is shown in the top of the 2nd column. Each UDG's ID number of in this catalog is shown in the bottom right of each cutout.}
    \label{fig:sbSample}
\end{figure}

UDGs are roughly on a red sequence in the Coma cluster \citep{Koda2015, Yagi2016}.
With the HSC color, they have (\textit{g}$-$\textit{r})$_{\rm HSC} \sim 0.55\pm 0.20$ (see also Section \ref{sec:color}).
The cosmological dimming correction may or may not be applied, depending on the studies.
Given these, the often-used threshold of $\mu_{\rm 0, g} = 24.0$ in $g$-band translates to $\langle \mu_{\rm r} \rangle_{\rm eff} =$ 24.4-24.8 in $r$-band (plus the cosmological dimming term when it is applied). 
Hence, our threshold of $\geq 24.1$ at $z=0.023$ is about 0.3 brighter than
the often-used threshold.  
Figure \ref{fig:sbSample} shows examples of UDGs organized by $\langle \mu_{\rm r} \rangle_{\rm eff}$.

Criterion 3:
The color cut is imposed to remove background objects, the objects too red for galaxies in the Coma cluster. Among the objects redder than \textit{g}$-$\textit{r}$=1$, 54 have SDSS spectra, and all of them are at higher redshifts ($z \geq 0.421$).
% we impose a redshift cut after criterion 3, removes 6

We measure the colors from the images themselves within a fixed aperture. The aperture color is measured using the \textit{g}- and \textit{r}-band fluxes of the images within a 2 kpc radius ($2 \times r_{\rm eff, r}$ of our smallest UDGs), excluding masked regions and the inner 1$\arcsec$ to avoid any potential nuclei. 

Applied individually, criterion 1, 2, and 3 removes 2600, 2553, and 486 objects respectively; applied sequentially, they remove 2600, 804, and 44 objects. 
All together they remove 3448 objects and leave 1504 objects.

\begin{figure}[htp]
    \centering
        \includegraphics[width=8cm]{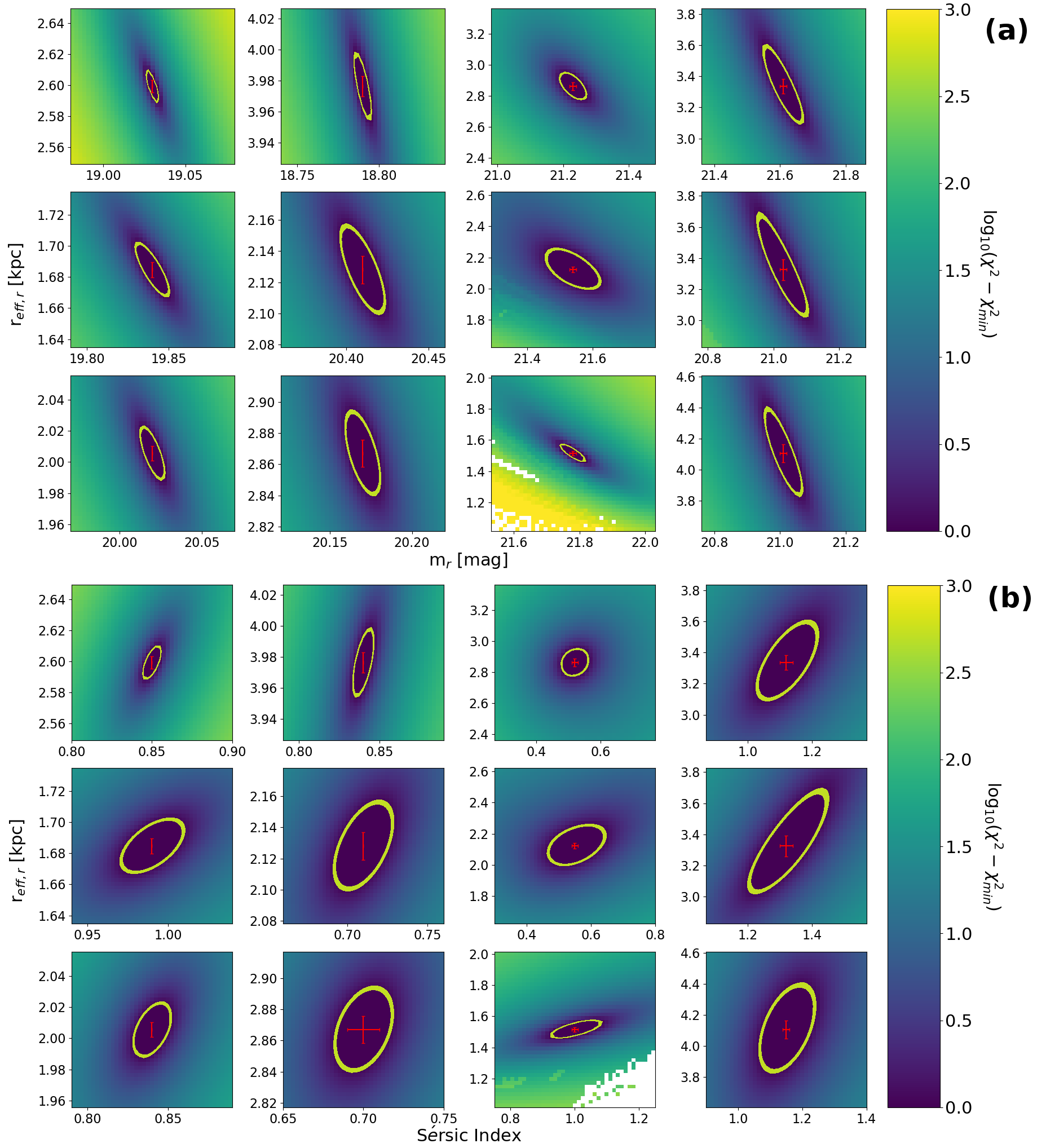}
        \caption{Change in $\chi^2$ from the best fit solution, $\chi^2_{\rm min}$, for the sample of UDGs in Figure \ref{fig:sbSample}. From the three-dimensional parameter space of $r_{\rm eff, r}$, $m_{\rm r}$, and $n_{\rm r}$, we show two projections: (a) $r_{\rm eff, r}$ and $m_{\rm r}$, and (b) $r_{\rm eff, r}$ and $n_{\rm r}$. The $\chi^2$ values are optimized in the other parameters for each fixed pair ($n_{\rm r}$, $r_{\rm eff, r}$) and ($m_{\rm r}$, $r_{\rm eff, r}$). The $\chi^2 - \chi^2_{\rm min} = 1$ contour is shown in yellow, while the error output from \textsc{GALFIT} is shown in red. Blank regions are where no convergent solution was found. As in Figure \ref{fig:sbSample}, columns are arranged according to $\langle \mu_{\rm r} \rangle_{\rm eff}$: from left to right, $\langle \mu_{\rm r} \rangle_{\rm eff}$ fainter than 24, 25, 26, and 27 mag$\cdot$arcsec$^{-2}$. The plot ranges span ($\Delta n$, $\Delta r_{\rm eff,r}$, $\Delta m_{\rm r}$) = (0.1, 0.1~kpc, 0.1~mag) for the left 2 columns, and (0.5, 1.0~kpc, 0.5~mag) for the right 2 columns.}
    \label{fig:chi2basin}
\end{figure}

By trial and error, we found that systematic errors stemming from the masking of contaminants and the sky-subtraction play significant roles in the fit (discussed in Section \ref{subsec:yagiCat}). In addition, the errors reported by \textsc{GALFIT} appear optimistic for the quality of the fits. For example, we take the 12 UDGs in Figure \ref{fig:sbSample} (arranged from brightest to faintest $\langle \mu_{\rm r} \rangle_{\rm eff}$ and examine the $\chi^2$ values in the parameter space of $r_{\rm eff, r}$, $n_{\rm r}$, and $r$. In this multidimensional space, an error surface (e.g., the surface of $\chi^2-\chi^2_{\rm min} = 1$) forms a manifold, and Figure \ref{fig:chi2basin} shows its two projections: (a) one derived with $r_{\rm eff, r}$ and $r$ held constant in each fit and the rest of the parameters (S\'ersic index, center position, axis ratio, position angle) free to be optimized, and (b) the other with $r_{\rm eff, r}$ and $n_{\rm r}$ fixed and the rest free. The sampling step is 1/41 of the widths. In Figure \ref{fig:chi2basin}a, the $\chi^2-\chi^2_{\rm min} = 1$ contours extend over the widths of about (0.03, 0.05, 0.2, 0.5) kpc and (0.02, 0.03, 0.10, 0.15) mag from the higher (the left column) to lower (right) $\langle \mu_{\rm r} \rangle_{\rm eff}$. In Figure \ref{fig:chi2basin}b, the contours extend over about (0.02, 0.03, 0.16, 0.22) in $n_{\rm r}$. The red crosses in each panel show the errors from \textsc{GALFIT}, which are smaller than the sizes of the $\chi^2-\chi^2_{\rm min} = 1$ contours. Correlations among the errors in $r_{\rm eff, r}$, $n_{\rm r}$, and $r$ are evident, especially for the fainter UDGs.

\subsection{Removal of Redshift Outlier}
\label{sec:zoutlier}
To remove objects which are already known to be outside the Coma cluster, we searched NED for spectroscopic redshifts
within a 3” radius of the 1504 UDG candidates.
We found redshift measurements for 38 objects.
With the adopted Coma cluster redshift range of z=[0.015, 0.035],
one of the 38 objects is at a lower redshift of z=0.0064 (SDSS DR7), 
and none are beyond $z>0.035$.

This leaves 1503 objects with $r_{\rm eff, r} \ge 1.0$~kpc and 774 with $>1.5$~kpc.
While our selection is based on the measurements in r-band, 535 of the 1503 satisfy
the traditional definition of $r_{\rm eff, g} = 1.5$~ kpc and $\mu_{\rm 0, g} \ge 24.0\,\rm mag/arcsec^2$ in g-band.

\subsection{Final Parameter Refinement}
\label{sec:finalRefinement}

While several of our UDGs are fit by a single S\'ersic profile that produces a flat \textsc{GALFIT} residual image, this is not always the case. In particular, several UDGs appear to be nucleated \citep{Yagi2016}. To account for these cases and determine the nucleation fraction for this catalog, we refit each UDG with a S\'ersic + PSF composite model, using the result from the single S\'ersic fit as the initial guess. The PSF component is also positioned initially at the center of the single S\'ersic fit. We ran \textsc{GALFIT} multiple times with a range of initial guesses for the PSF magnitude ($m_{\rm psf}$), from 17 to 27 in steps of 0.25. The fits are made with no parameters fixed, and both the S\'ersic and PSF positions may drift independently from the initial guess and from each other. We pick the result with the lowest $\chi^2$. In cases where no convergent S\'ersic + PSF fit was found, the UDG is considered non-nucleated. 

We set two criteria for classifying nucleated UDGs (i.e., choosing the S\'ersic + PSF results over the single S\'ersic ones). First, the S\'ersic and PSF components must be centered within one arcsecond of each other. Second, we compare the fit quality of the best S\'ersic + PSF fit to the single S\'ersic fit. To objectively compare the two models, and to take into account the difference in the number of free parameters, we use the Bayesian information criterion (BIC) \citep{Schwarz1978},

\begin{equation}
    BIC = \chi^2 + n_{\rm param} \ln(N),
\end{equation}

where $N = n_{\rm dof} + n_{\rm param}$ is the number of pixels used in the fit. The fit with the lowest BIC is considered a better fit result. With this comparison, 309 galaxies are classified as nucleated, 183 of which are UDGs with $r_{\rm eff}\geq 1.5$ and 126 are sub-UDGs with $r_{\rm eff}=1.0$-$1.5$.

\subsection{Note on Contamination by Galactic Cirrus}
\citet{Zaritsky2021} checked probable cirrus contamination in the \textit{SMUDGes} catalog by checking cirrus emission in \textsc{WISE} 12 $\mu$m ($\geq 0.1$ MJy/sr) \citep{MeisnerFinkbeiner2014} and \textit{Planck} $\tau_{\rm 353}$ ($\geq 0.05$) images \citep{Planck2014, Green2018}. They found that only 1.8\% of their UDGs were misclassified cirrus. Applying the same method to our UDGs, we did not find any UDG to be contaminated by cirrus emission. We note that our visual inspection during the UDG selection did not identify irregular shapes expected for cirrus emission.

\section{Catalog}
\label{sec:Catalog}

\begin{figure*}[htp]
    \centering
    \includegraphics[width=18cm]{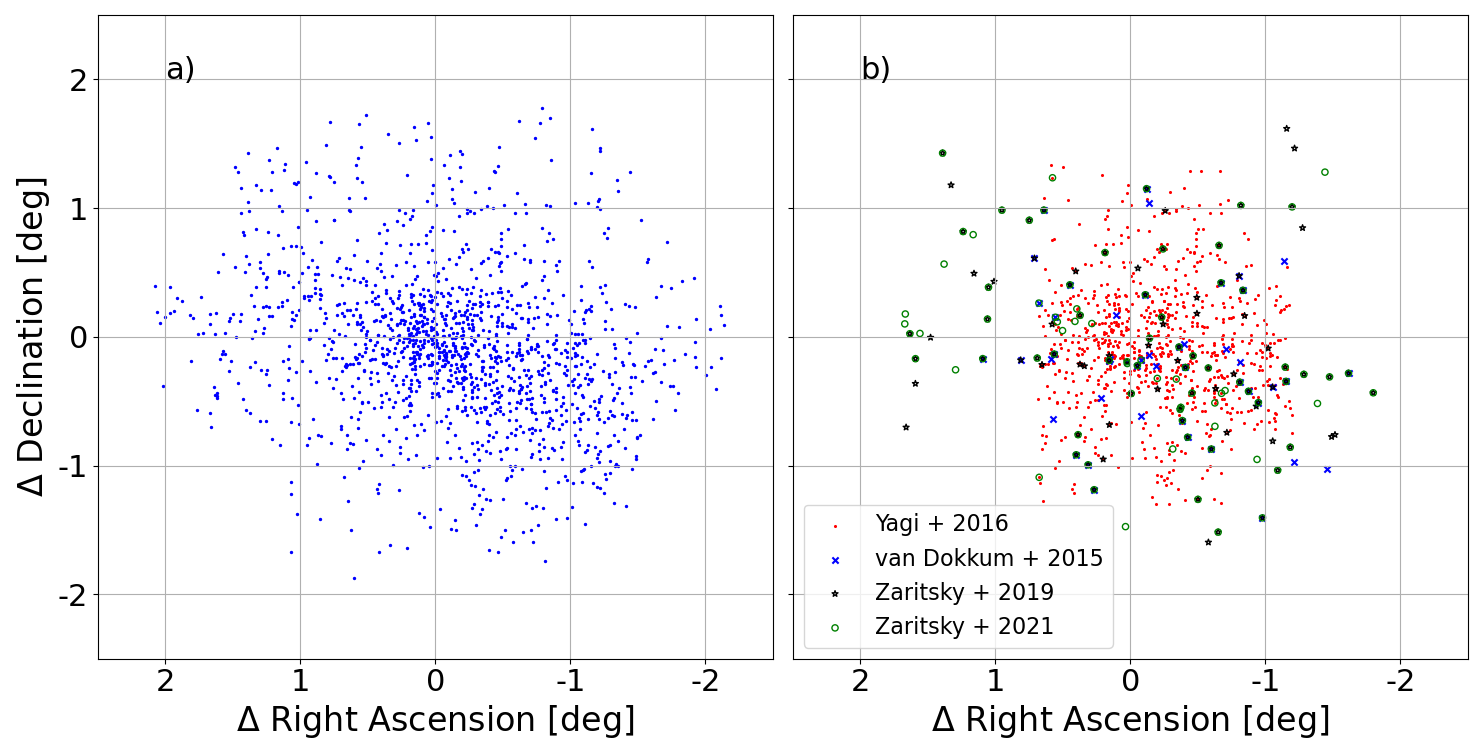}
    \caption{a) The spatial distribution of UDGs around the Coma cluster center $(\alpha_{\rm J2000}, \delta_{\rm J2000}) =$ (12:59:42.8, +27:58:14). b) UDGs from other catalogs in the same area.}
    \label{fig:radecScatter}
\end{figure*}

The catalog of best-fit parameters from \textsc{GALFIT} in both \textit{g}- and \textit{r}-band, along with their aperture colors, is given in Table 1 %\ref{tab:catalog}
(the table in its entirety, including fit errors from \textsc{GALFIT} is available in machine-readable form). The right ascension, declination, $m$, $r_{\rm eff}$, $n$, $q$, $PA$ were taken from the output of \textsc{GALFIT}. The $PA$ is defined such that $PA=0$ when the major axis lies along the north, increasing counterclockwise. The \textit{g}-band positions and S\'ersic indices are set by the corresponding \textit{r}-band fit values. Other parameters are fit independently in \textit{g}- and \textit{r}-bands. For cases where the UDG is nucleated, $m_{\rm psf}$ is also listed. Figure \ref{fig:radecScatter} shows the distribution of UDGs in this catalog compared to 3 other UDG catalogs in the same area \citep{PVD2015, Yagi2016, Zaritsky2019}. 
We calculate both $\langle \mu \rangle_{\rm eff}$ and $\mu_{\rm 0}$ from the \textsc{GALFIT} parameters as:

\begin{equation} \label{eq3}
    \begin{split}
        \langle \mu \rangle_{\rm eff} & = m + 2.5\log_{\rm 10}\left( 2\pi q r_{\rm eff}^{2}\right),\\
        \mu_{\rm 0} & = \langle \mu \rangle_{\rm eff} + 2.5\log_{\rm 10}\left( \frac{n}{b^{2n}}\Gamma(2n)\right)
    \end{split}
\end{equation}
where $\quad 2\gamma(2n, b) = \Gamma(2n)$.
$\Gamma$ and $\gamma$ are the complete and incomplete gamma functions respectively \citep{Graham2005}.
%where $b$ is calculated using the expression frovided in \citet{Ciotti1999}.

In the following subsections, we will compare our new catalog with other catalogs in the literature. 
In particular, we empirically evaluate errors in our own measurements in comparison with the previous Subaru catalog (Section \ref{subsec:yagiCat}). 
In summary, 
we estimate the random errors to be (0.15~mag, 0.16~kpc, 0.13) in ($m$, $r_{\rm eff}$, $n$).
The systematic errors, likely due to errors in sky subtraction, can be (0.04~mag, 0.05~kpc, 0.09). 

\subsection{Previous catalogs of Coma UDGs}
\label{sec:UDGCatalog}

We compare the parameters in this and previous catalogs. As we will see, the measurements of UDGs are inherently difficult and suffer significantly from random and systematic errors due to different sensitivities, resolutions, and artifacts. Therefore, the selection criteria have to be optimized to realistically reflect each data quality. This is part of the motivation for having a tolerance in our catalog
(see Sections \ref{sec:term} and \ref{sec:galSelection}) and including sub-UDGs.

\subsubsection{Comparison with Yagi et al.(2016)}
\label{subsec:yagiCat}
\citet{Yagi2016} used \textit{R}-band Suprime-Cam images to find 854 UDGs in a smaller, approximately 1.7$\arcdeg$ $\times$ 2.7$\arcdeg$ area around the center and the western half of the cluster, supplemented with \textit{B}-band information from \citet{Yamanoi2012} for 232 UDGs in a sub-region within the \textit{R}-band area. Compared to these Suprime-Cam UDGs (``SC-UDGs"), we have approximately doubled the searched area to a roughly 2$\arcdeg$ radius around the cluster center, and added color information to all UDGs. The number of objects has nearly doubled from 854 SC-UDGs (in the limited area) to 1503 (across the full cluster), 774 of which have $r_{\rm eff, r} \geq 1.5$.

\begin{figure}[htp]
    \centering
    \includegraphics[width=8cm]{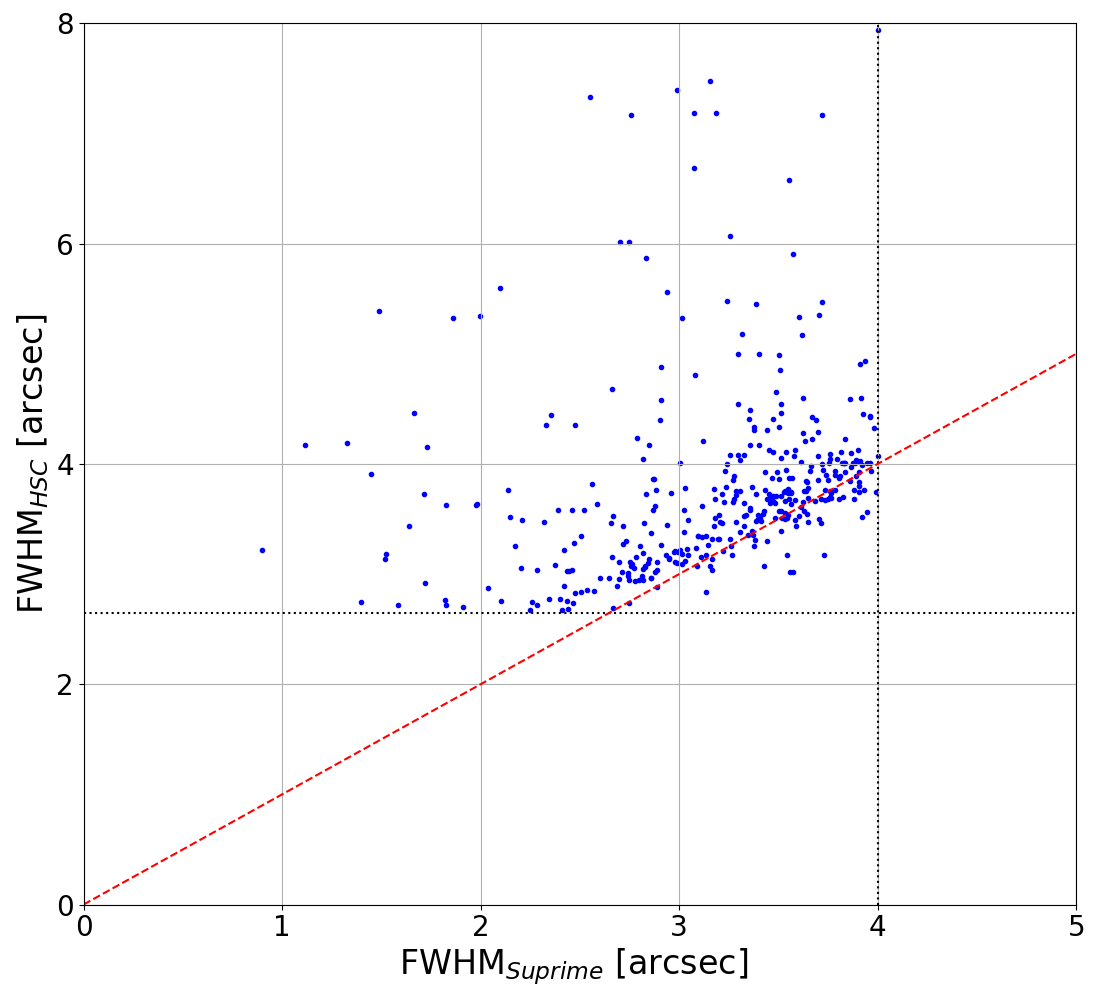}
    \caption{Correlation between the \textsc{SExtractor} FWHM between Suprime-Cam and HSC for the UDGs in the area of \citet{Yagi2016} that were rejected by FWHM in our analysis. Our selection limits are the blacked dotted lines, while the red dashed line has unit slope. The FWHM are generally measured as smaller in the Suprime-Cam images.}
    \label{fig:fwhmComparison}
\end{figure}

Within the Suprime-Cam search area, there are 1059 UDGs, 674 of which are in \citet{Yagi2016}, and the remaining 385 are new UDGs. 
By looking back the intermediate steps in \citet{Yagi2016}, $\sim$75\% of the 385 new UDGs were detected in their \textsc{SExtractor} run, but rejected by their selection criterion of FWHM$> 4\arcsec$ with \textsc{SExtractor}.
Figure \ref{fig:fwhmComparison} shows that the FWHMs from Suprime-Cam are smaller than those from the new HSC.
The remaining $\sim$25\% of the 385 were not detected by their \textsc{SExtractor} run
mainly due to blending and bleeding.
With the cleaning algorithm, we are able to detect them properly.

Of all the 854 SC-UDGs, 674 are retained in the new catalog. The remaining 180 are rejected, most of which are now measured to be smaller than the size cutoffs in our analysis.

\begin{figure*}[htp]
    \centering
    \includegraphics[width=18cm]{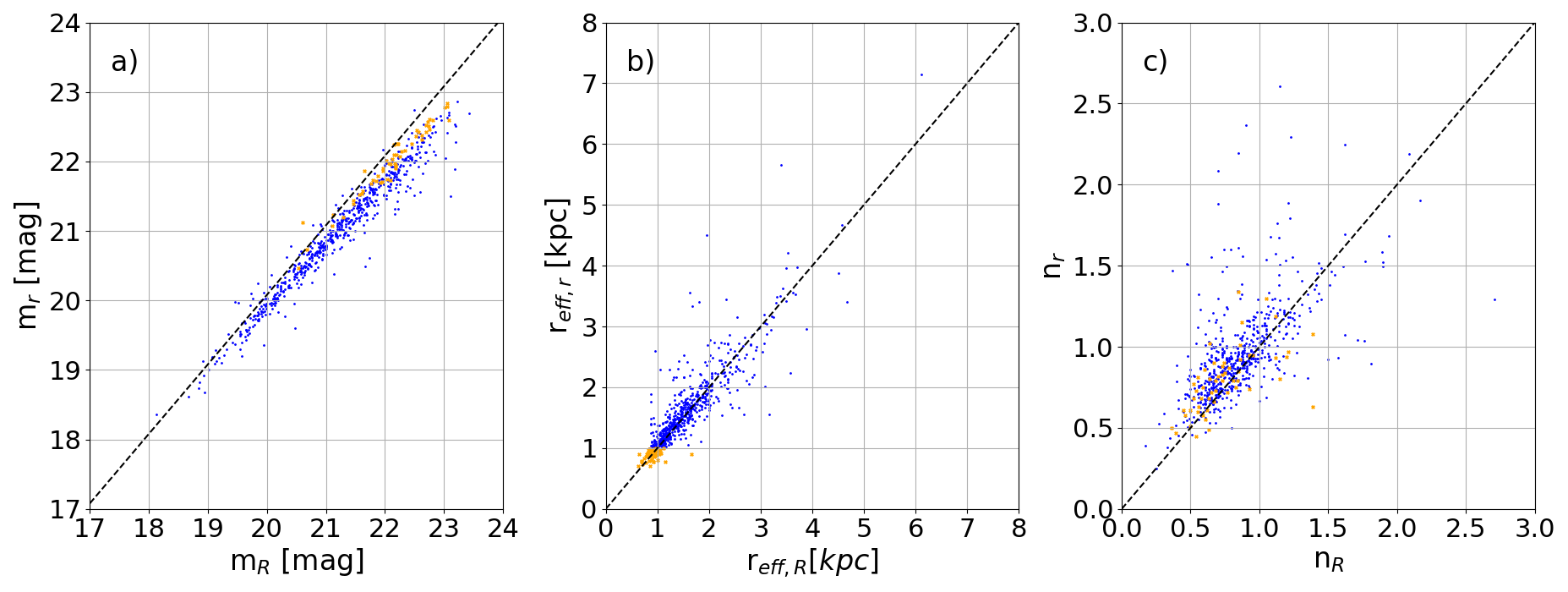}
    \caption{Comparison of \textit{r}-band values to corresponding \textit{R}-band values in \citet{Yagi2016}. UDGs in this work are plotted as blue dots, while the 65 SC-UDGs that were removed by the \textsc{GALFIT} $r_{\rm eff, r}$ selection criterion are shown as orange crosses. For $m$, the dashed line is equation (\ref{eqR}), assuming the median color \textit{g}-\textit{r} = 0.55.}
    \label{fig:new-yagi_rComparison}
\end{figure*}

Figure \ref{fig:new-yagi_rComparison} compares the Suprime-Cam \textit{R}-band structural parameters with the corresponding HSC \textit{r}-band values. The 2 sets of parameters are correlated, ensuring the general consistency between the previous study and ours using the same telescope.
The systematic offsets of only 0.04 kpc and 0.07 in $r_{\rm eff}$ and $n$ are much smaller than their scatters of 0.34 kpc and 0.25, respectively. The comparison of magnitude shows a small scatter of 0.15, but also shows a non-negligible systematic offset as a function of $m$.
The offset becomes larger, $\Delta (R-r) \sim -0.21$,
when the color conversion ($r$-$R=0.07$) is taken into account.

The most likely cause of this offset is an over-subtraction of the sky in the Suprime-Cam study due to the difficulty of determining the extents of UDGs' faint outskirts. 
In fact, the offset diminishes if we re-evaluate the sky levels 
as ``2.5$\times$median - 1.5$\times$mean" in the cutout images of Suprime-Cam (the same method we used for the new data; see Section \ref{subsec:localSky}).

\begin{figure*}[htp]
    \centering
    \includegraphics[width=18cm]{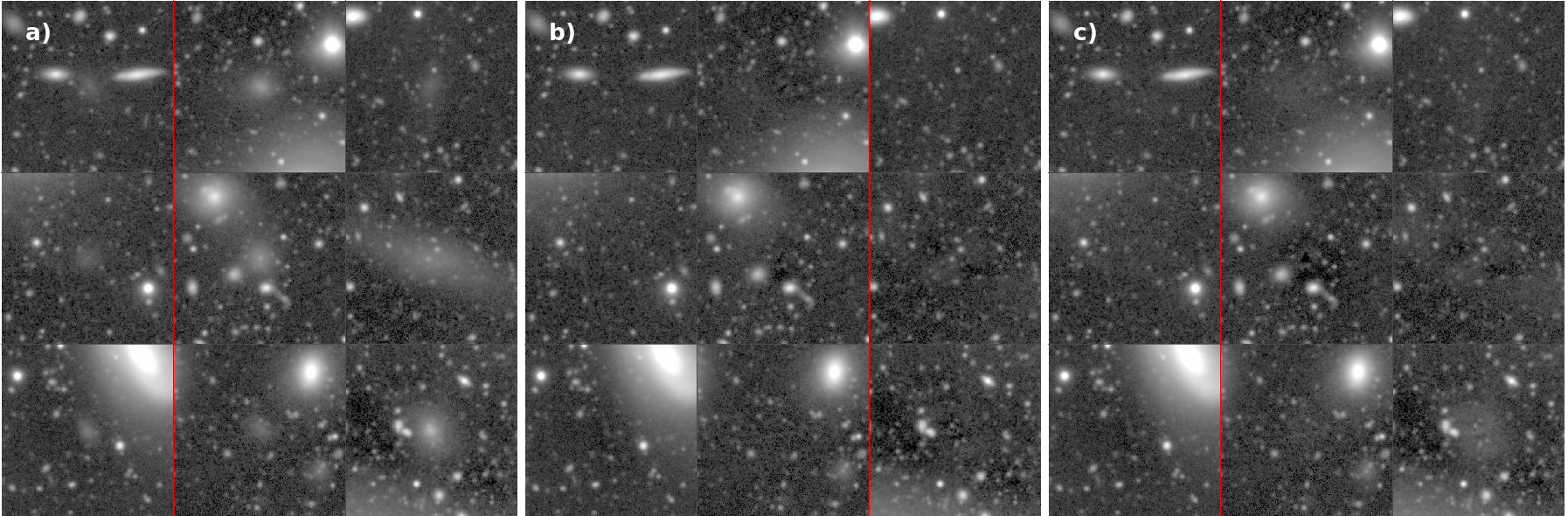}
    \caption{Nine sample UDGs that are outliers in the correlation plot (Figure \ref{fig:new-yagi_rComparison}). (a) \textit{r}-band HSC image. (b) Residual with best fit \textit{r}-band values in this work. (c) Residual with \textit{R}-band fit values from \citet{Yagi2016}. The $m_{\rm R}$ for the S\'ersic model is converted to the equivalent $m_{\rm r}$ using equation (\ref{eqR}). Comparing the 2 sets of residuals, we find the different fit values produce little difference in the quality of the residuals.}
    \label{fig:new-yagi_outliers}
\end{figure*}

This level of discrepancies among measurements may not be avoidable when the objects are as faint as UDGs. 
A representative set of 9 objects that deviate by 3$\sigma$ from the solid lines in Figure \ref{fig:new-yagi_rComparison} is shown in Figure \ref{fig:new-yagi_outliers}a. Figures \ref{fig:new-yagi_outliers}b and c show the residual images when subtracting a S\'ersic profile with the best fit \textit{r}-band parameters found in this work, and with the \textit{R}-band parameters from \citet{Yagi2016} respectively. We use equation (\ref{eqR}) for the color conversion. The differences between most of the residuals are subtle, despite the fit values being outliers in the correlation plot.

\begin{figure}[htp]
    \centering
    \includegraphics[width=8cm]{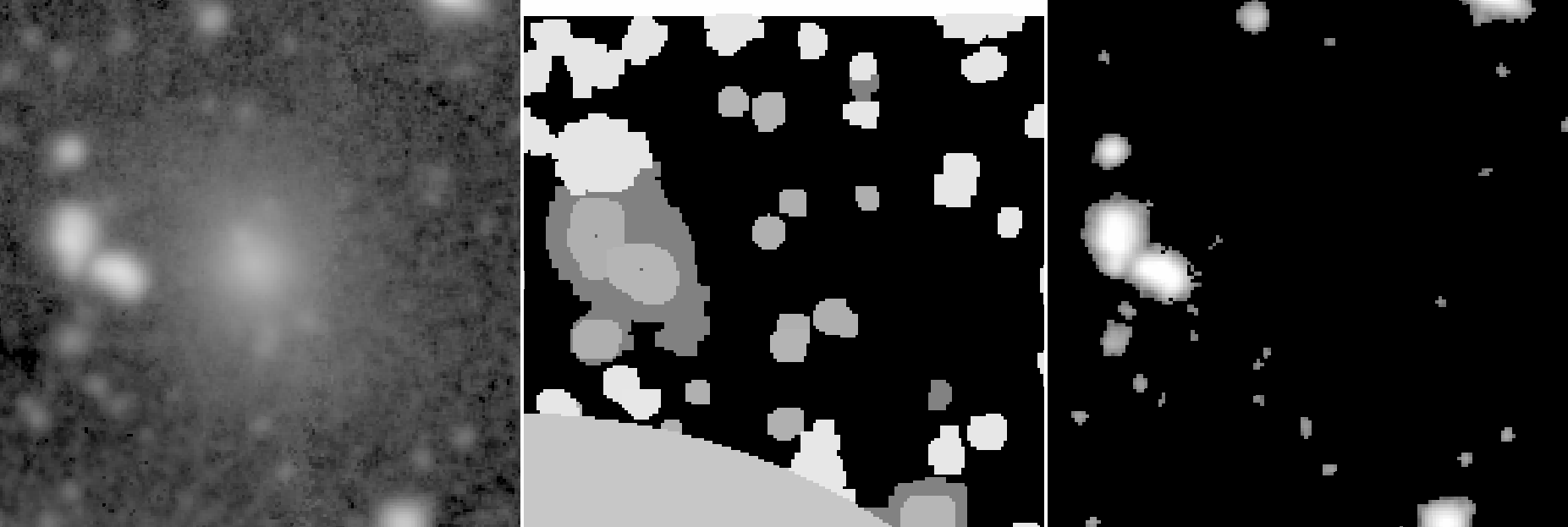}
    \caption{Comparison of contaminant masks between this work and \citet{Yagi2016} for one outlier UDG (bottom right in Figure \ref{fig:new-yagi_outliers}). From left to right: zoomed-in \textit{r}-band HSC raw image, contaminant mask generated in this work, and contaminant mask from \citet{Yagi2016}.}
    \label{fig:hscSCmask}
\end{figure}

More extreme outliers from the correlations may stem from a difference in the masking of contaminants. In Figure \ref{fig:hscSCmask}, we compare the masks between this work and \citet{Yagi2016} for one of the outlier UDGs in Figure \ref{fig:new-yagi_outliers}. The new masking technique presented in Section \ref{sec:removecompact} identifies and masks point sources on top of and around the UDG efficiently (the impacts of insufficient masking on top of UDGs is discussed again in Section \ref{subsec:smallObj}). As shown in Figure \ref{fig:chi2basin}, fainter UDGs have wider error bars, and their parameters are more difficult to constrain. This is reflected by the increase in scatter towards fainter $m$ in Figure \ref{fig:new-yagi_rComparison}a. While the Suprime-Cam and new HSC catalogs are mostly consistent, these discrepancies demonstrate the difficulty of the UDG fit. The UDGs are intrinsically faint, which limits the accuracy of any catalog, including the one presented in this paper.

The comparisons of the two measurements with the similar data offer an opportunity to estimate errors in the UDG measurements (see Figure \ref{fig:new-yagi_rComparison}).
As discussed above, the differences between the two arise from several sources of errors: including, but likely not limited to, (a) the different data realizations, (b) different sky subtractions, (c) different masks, and (d) fitting errors with GALFIT.
It is difficult to separate them completely, but we can have a sense.
The scatters between the two measurements likely indicate random errors, due to the combination of (a) and (d).
These random errors are (0.15~mag, 0.16~kpc, 0.13) in ($m$, $r_{\rm eff}$, $n$),
where we used only the 90\% data around the center of each parameter distribution to exclude outliers, which are due to (c).
We take these random errors as the errors in our current measurements,
but if the errors in the two measurements contribute equally, we should divide the random error values by $\sqrt{2}$.
Errors due to (c) occur only occasionally, but when they occur, they are large.
It is difficult to characterize these errors by one number,
but from Figure \ref{fig:new-yagi_rComparison},
the most extreme deviations are $\sim 3$~mag, $2.5\kpc$, and 1.5 in $n$.
The systematic shifts between the two measurements are one realization of the systematic error due to (b).
The amounts are (0.21~mag, 0.06~kpc, 0.09). 

\subsubsection{Comparison with van Dokkum et al. (2015)}
\label{subsec:dfCat}
All 47 UDGs detected with the Dragonfly telescope (``DF-UDGs") from \citet{PVD2015} are detected in this catalog\footnote{Note that DF27 was matched using the coordinates in \citet{Yagi2016}, which are offset by about 15" from the original coordinates in \citet{PVD2015}}. \citet{Yagi2016} intentionally adjusted the selection criteria to include DF-UDGs as a fiducial set. 

As discussed in Sections \ref{sec:removecompact} and \ref{sec:remblend}, the confusion of foreground and background objects is a major obstacle in identification of UDGs at the distance of the Coma cluster. This study found a much larger number of UDG mainly due to the higher resolution being able to separate contaminants from the galaxies. For \textsc{GALFIT} fitting, \citet{PVD2015} used CFHT \textit{g}- and \textit{r}-band images, and assumed a pure exponential ($n$ = 1) profile fitting to increase the stability of the fit. We leave $n$ as a free parameter for all galaxies, due to the high surface brightness sensitivity. The $r_{\rm eff}$ are in agreement without systematic offsets despite the difference in band (\textit{g}-band and \textit{r}-band), but as a reminder, the \textit{g}-band S\'ersic index is held fixed to the \textit{r}-band value. 

2 of the 47 DF-UDGs do not have $r_{\rm eff, r} \geq 1.5$, and are sub-UDGs. These 2 DF-UDGs have an $r_{\rm eff, g}$ of exactly 1.5 in \citet{PVD2015}, so these are marginal cases whose sizes hover about the cutoff depending on band, fit quality, and sky estimation. Such marginal cases are the impetus for using a lower $r_{\rm eff}$ selection cut. The remaining 45 DF-UDGs  have $r_{\rm eff, r} \geq 1.5$, so the majority of the DF-UDGs are on the larger half of this catalog. 

It is not straightforward to assess the completeness of the \citet{PVD2015} study with respect to the new catalog.
\citet{PVD2015} did not attempt completeness, which already explains the large difference in the numbers of detections. While the selection criterion of (a) $r_{\rm eff}\geq 1.5$ and (b) $\mu_{\rm 0} \geq 24$ are often referred to \citet{PVD2015}, their selection was not based on these criteria, but on a set of others tuned for \textsc{SExtractor} outputs on the images of 6" resolution.
Our detections are made on a substantially different basis in image quality (e.g., 1" vs 6" PSF size and sensitivity), band for detection (e.g., \textit{r} vs \textit{g}+\textit{r} combined), and software for measurement (\textsc{GALFIT} vs \textsc{SExtractor}).
As the best effort, however, we adopt (a), (b), and two additional criteria that mimic \citet{PVD2015}'s:
(c) $20 \leq m_{\rm g} \leq 23$,
and (d) low central concentration $|m_{\rm g} -m_{\rm aperture}| \leq 1.8$ with 3$\arcsec$-radius aperture magnitude $m_{\rm aperture}$ in \textit{g}-band.
By applying (a)-(d) to our catalog, we yield 687 objects.
Of these, 33 are also found in \citet{PVD2015}.

\begin{figure*}[htp]
    \centering
    \includegraphics[width=18cm]{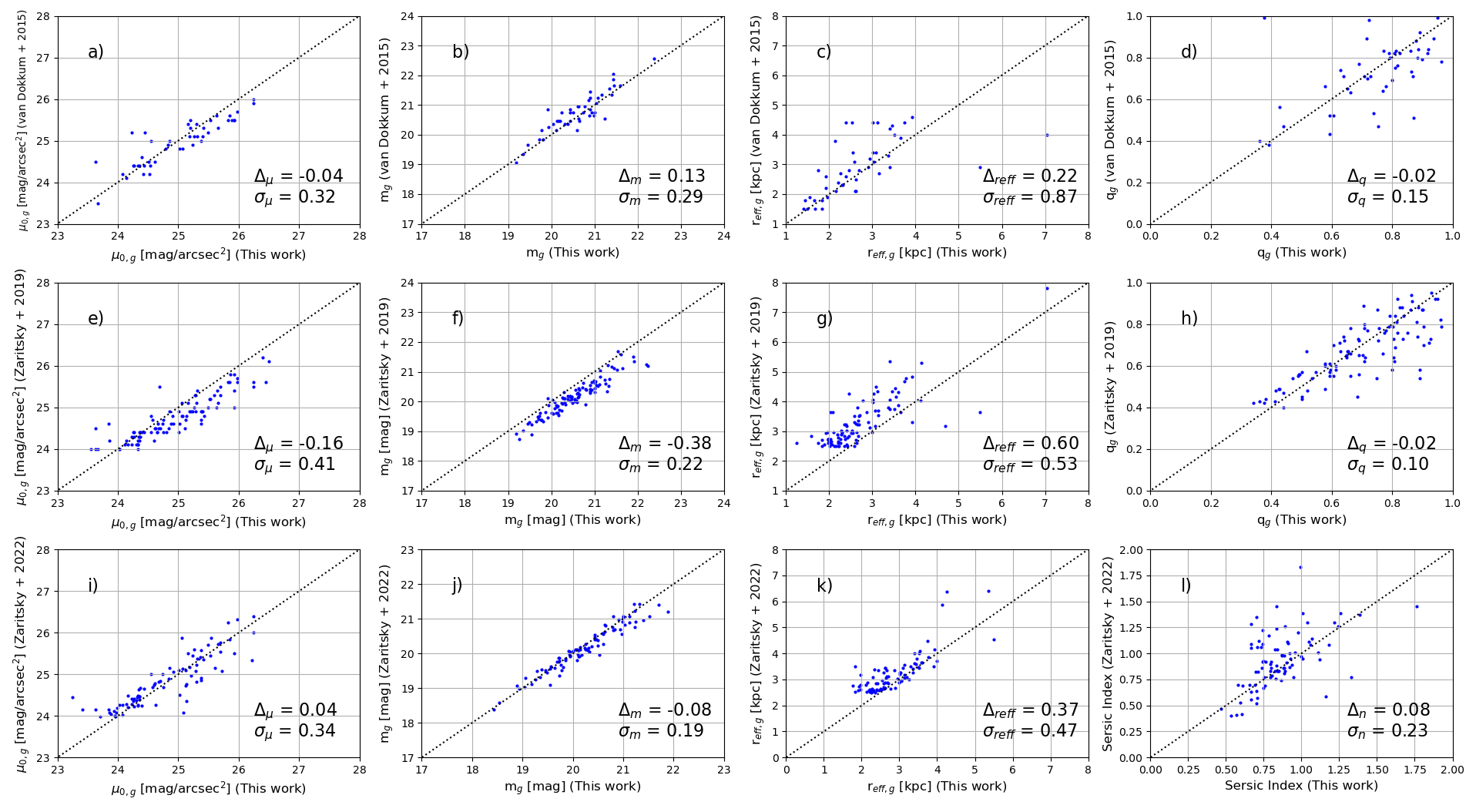}
    \caption{Comparison of $\mu_{\rm 0, g}$, $m_{\rm g}$, $r_{\rm eff, g}$, and $q_{\rm g}$ (or $n$) between this work and (a)-(d) the 47 UDGs in \citet{PVD2015}, (e)-(h) the 98 UDGs in \citet{Zaritsky2019}, and (i)-(l) the 86 UDGs in \citet{Zaritsky2022}. (l) compares n as it is set free in \citet{Zaritsky2022}. Meanwhile, since n is fixed to 1 in \citet{PVD2015} and \citet{Zaritsky2019}, (d) and (h) compare q instead.} The dashed lines have a unit slope. Systematic offsets are apparent between the parameters of this work and \citet{Zaritsky2019}.
    \label{fig:catalogCompScatter}
\end{figure*}

Figures \ref{fig:catalogCompScatter}a-d compare \textsc{GALFIT} measured \textit{g}-band parameters between \citet{PVD2015} and ours for the 47 DF-UDGs. \citet{PVD2015} used the archival CFHT data for the measurements.
\citet{PVD2015} converted the CFHT to SDSS photometry.
Overall, the two are consistent.
The small offsets in $\mu_{\rm 0}$ are expected from the band difference between SDSS and HSC ($\sim$ 0.1 mag). Given that, the offsets are smaller than the scatters.
The $r_{\rm eff}$ are also consistent, and the large standard deviation of 0.87 is larger than the systematic offsets of 0.22. We note that between this work and \citet{PVD2015}, there is a negligible difference in both distance modulus (35.05 vs 35.057 mag respectively) and scale (0.473 vs 0.475 kpc$\cdot$arcsec$^{-1}$ respectively).

\subsubsection{Comparison with Zaritsky et al. (2019)}
\label{subsec:zaritskyCat}

\begin{figure}[htp]
    \centering
    \includegraphics[width=5cm]{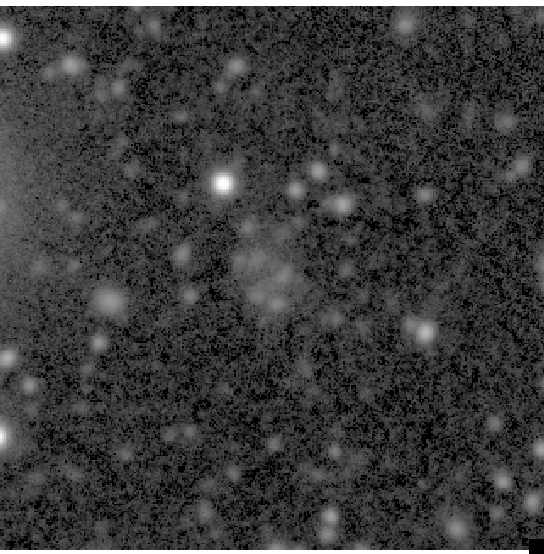}
    \caption{\textit{r}-band HSC image of SMDG 1306050+273627. It is cataloged in \citet{Zaritsky2019}, while rejected in this work.}
    \label{fig:missingSmudges}
\end{figure}

 \citet{Zaritsky2019}, the first result of the \textit{SMUDGes} survey, covers a much wider area beyond the Coma cluster (approximately 300 square degrees, which is 20 times larger than our area), but with shallower depth. 99 \textit{SMUDGes} UDGs are present in our field coverage, with 98 having corresponding entries in our catalog. \textit{SMUDGes} includes only objects with $r_{\rm eff} \geq 2.5$ in the \textit{g}-band. The object (SMDG 1306050+273627), rejected in our work, is shown in Figure \ref{fig:missingSmudges}. Our selection procedure rejected this object as a blended star cluster.

Of the 98 \textit{SMUDGes} UDGs, 51 do not fulfill their own selection criteria with our measured parameters. 47 of them have $r_{\rm eff, g} \leq 2.5$, and 7 have $\mu_{\rm 0, g} < 24.0$. Figures \ref{fig:catalogCompScatter}e-h show the comparison of their \textit{g}-band $\mu_{\rm 0}$, $m$, $r_{\rm eff}$, and $q$. There are clear offsets. Our $\mu_{0}$ are on average fainter than those in \citet{Zaritsky2019} by 0.16, with a standard deviation of 0.41. On the other hand, our $m$ are fainter by 0.38 with a standard deviation of 0.22. Our $r_{\rm eff}$ are on average smaller by 0.60 with a standard deviation of 0.53. We note that between this work and \citet{Zaritsky2019}, there is a negligible difference in both distance modulus (35.05 vs 35.057 mag respectively) and scale (0.473 vs 0.475 kpc$\cdot$arcsec$^{-1}$ respectively).

Pinning down the exact cause of the offsets is difficult without their data. 
However, we found that adding a constant positive background can reduce the offsets. Adding a constant sky of 28.5 mag$\cdot$arcsec$^{-2}$ to the UDG cutouts makes the offsets of all the parameters smaller than the scatters in Figure \ref{fig:catalogCompScatter}. 
Since we found the same course of errors between the Suprime-Cam and HSC studies with the same telescope (Section \ref{subsec:yagiCat}), it would not be a surprise if the sky-subtraction is again an issue here.

The other possibilities include differences in the masks of compact objects
and in the treatment of the S\'ersic index in fitting.
\citet{Zaritsky2019} used wavelet transformations,
as opposed to the unsharp masking (Sections \ref{sec:removecompact} and \ref{subsec:smallObj}).
They also fixed $n=1$, while we set it free.
Our fitting found that the average S\'ersic index of the 98 \textit{SMUDGes} UDGs is 0.86, and the difference from $n=1$ could alter the sizes and $\mu_{\rm 0}$ (see equation \ref{eq3}).

\begin{figure}[htp]
    \centering
    \includegraphics[width=8cm]{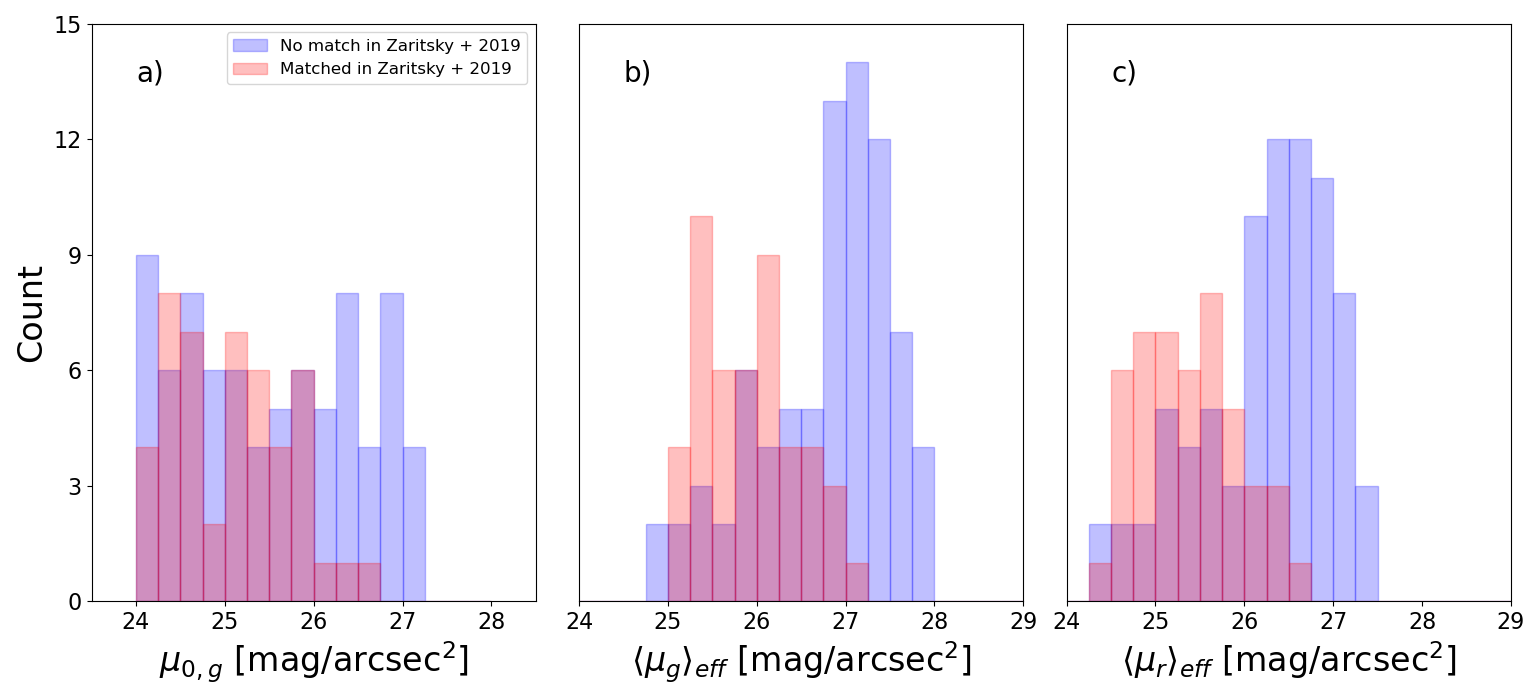}
    \caption{Comparisons of the objects matched with \citet{Zaritsky2019} (47, red) and unmatched (79, blue) UDGs that satisfy the \citet{Zaritsky2019} selection criteria: (a) $\mu_{\rm 0, g}$, (b) $\langle \mu_{\rm g} \rangle_{\rm eff}$, and (c) $\langle \mu_{\rm r} \rangle_{\rm eff}$. The unmatched UDGs are generally fainter in all of the surface brightnesses. }
    \label{fig:missingSmudgesSB}
\end{figure}

If we apply the selection criteria of \textit{SMUDGes} ($r_{\rm eff, g} \geq 2.5$ and $\mu_{\rm 0, g} \geq 24.0$) to our measurements of the new catalog, we find 126 objects.
Of these, only 47 are found in \textit{SMUDGes} \citep{Zaritsky2019}, and the rest 79 are not.
Figure \ref{fig:missingSmudgesSB} shows the $\langle \mu \rangle_{\rm eff}$ and $\mu_{\rm 0}$ distributions of the 126 UDGs,
separating the 47 matched and 79 unmatched UDGs with \citet{Zaritsky2019}. Approximately half of the 79 unmatched UDGs are fainter than the 47 matched UDGs.

If we estimate the combined errors of \textit{SMUDGes} and ours by comparing the two measurements as we did in Section \ref{subsec:yagiCat}, the random errors are (0.22~mag, 0.53~kpc) in ($m$, and $r_{\rm eff}$). The systematic errors are (0.38~mag, 0.60~kpc) and are likely due to the errors in sky subtraction.
It is difficult to separate the error sources between the two measurements, however, these errors are larger than those found in the comparisons of the two Subaru measurements (Section \ref{subsec:yagiCat}). 

%\textbf{Note that in the above, we compared our results with the Coma cluster analysis in \textit{SMUDGes} \citep{Zaritsky2019}. Since then, \textit{SMUDGes} has extended the coverage substantially to $\sim$ 15,000 square degrees \citep{Zaritsky2022}. The S\'ersic index $n$ is also fit, rather than fixed to $n=1$. Recent \textit{SMUDGes} studies, while not in the Coma cluster, estimated errors by simulations, i.e., by placing model UDGs in sky images and by measuring their parameters \citep{Zaritsky2021,Zaritsky2022}. Detailed comparisons of their errors with ours are beyond the scope of this paper. However, if we naively take medians of their cataloged errors as ``typical" errors, their random errors are about (0.16, 0.23, and 0.10) in ($m$, $r_{\rm eff}$, and $n$) at the 15.1 and 84.9 percentiles, and the systematic errors are about (0.04, 0.05, and -0.02), respectively.For the errors in $r_{\rm eff}$, we calculated typical fractional errors and multiplied $r_{\rm eff} =$ 2.5 kpc, the radius at their selection threshold. These errors are smaller than our empirical estimations presented above. The simulations do not account for the errors in sky subtraction, which might explain this difference, since we found those errors as a major source of the errors.}

\subsubsection{Comparison with Zaritsky et al. (2022)}

%%% Z19 vs Z22
More recently, \textit{SMUDGes} has substantially extended the coverage to $\sim$ 15,000 square degrees  \citep{Zaritsky2022}. 
The procedure for UDG detection and measurements was improved from \citet{Zaritsky2019}.
For example, \citet{Zaritsky2022} measured all parameters with $n$ free,
while \citet{Zaritsky2019} measured most parameters with $n=1$ fixed, except magnitude ($m$)
which were determined without  $n$ fixed.
\citet{Zaritsky2022} found 88 \textit{SMUDGes} UDGs in the Coma cluster area that we covered,
of which only 58 were in \citet{Zaritsky2019}. 
On average, their new $r_{\rm eff}$ are smaller by 0.24~kpc, and the new $m$ are  fainter by 0.20 mag than the old ones.
The measured $n$ are $\sim 0.8$ rather than $1$, which may cause the difference in $r_{\rm eff}$.
However $m$ is measured with $n$ free in both studies, and hence, the difference in $m$ suggests an existence of an additional systematic difference independent of $n$.
One possibility, among others, 
could be sky-subtraction (see Section \ref{subsec:yagiCat}).

Of the 88 new \textit{SMUDGes} UDGs, 86 have a corresponding entry in our catalog. The remaining two are SMDG1304338+264623 and SMDG1252075+272654. 
The former is at z= 0.0064, not in the Coma cluster redshift (removed in Section \ref{sec:zoutlier}).  
The latter is rejected due to contamination by an optical ghost in our image.
Their selection criteria include $r_{\rm eff}>2.5$~kpc.
Of the 86, 57 have $r_{\rm eff}>2.5$~kpc in our catalog as well.

The parameters in the new \textit{SMUDGes} catalog are more consistent with ours. 
Figures \ref{fig:catalogCompScatter}i-l compare this work and \citet{Zaritsky2022}. 
For the objects with $r_{\rm eff}>2.5$~kpc in both catalogs,
the differences in ($\mu_{0}$, $m$, $r_{\rm eff}$, $n$)
are (0.01, -0.11, 0.47, 0.07) in average and (0.07, -0.04, 0.24, 0.00) in median.
These differences were calculated as their average/median values minus ours,
and we used their values after their model-based bias correction.
The differences are small except for $r_{\rm eff}$.
We note that the difference between \citet{Zaritsky2019} and \citet{Zaritsky2022}
were even larger, and by comparing the averages, the former were 0.60~kpc larger than the latter.

%\citet{Zaritsky2021} estimated errors by simulations, i.e., by measuring the parameters of model UDGs placed in sky images.
%They report random errors for each object (at the 68\% confidence interval) and systematic errors (the median difference of input and measured values). 

%If we naively take averages of the $1\sigma$ values as ``typical" random errors, they are about (0.17, 0.17, 9.2\%, and 0.10) in ($\mu_{0}$, $m$, $r_{\rm eff}$, and $n$). The systematic errors are about (0.06, 0.04, -2\%, and -0.02). These random errors may be an overestimation based on our comparisons between our catalog and theirs. Only $68\%$ of the 86 UDGs (i.e., the  overlaps in the twe catalogs) are expected to be in the $1\sigma$ range, but we find (79\%, 79\%, and 71\%) of the UDGs in ($m$, $r_{\rm eff}$, and $n$). We should note that our comparison of the two catalogs measured the combined errors of the two, while the errors estimated by \citet{Zaritsky2021} are are for their catalog alone. If their estimations were correct, we should have found less than 68\% within the $1\sigma$ range.

\subsection{Effect of small objects on parameters}
\label{subsec:smallObj}

\begin{figure*}[htp]
    \centering
    \includegraphics[width=18cm]{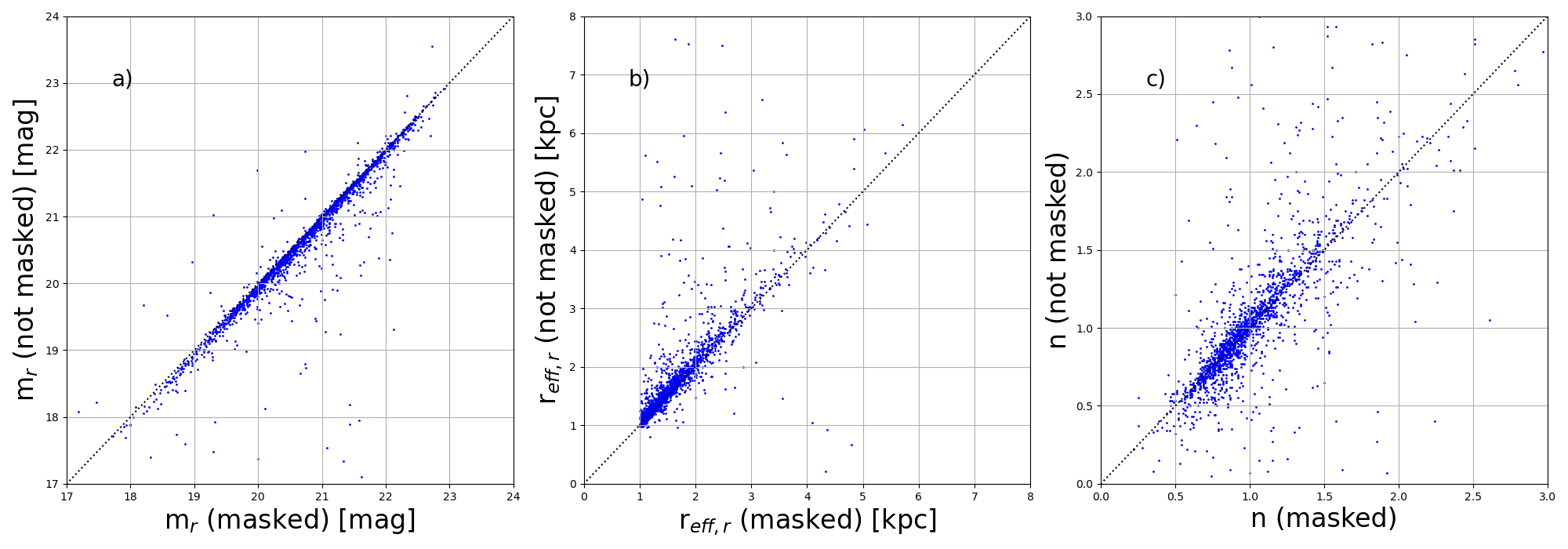}
    \caption{\rm Comparison of best-fit \textsc{GALFIT} \textit{r}-band (a) $m$, (b) $r_{\rm eff}$, and (c) $n$, when compact objects on top of the UDG are masked or not masked. The blacked dashed lines have unit slope. Some of the parameters derived without the compact object masks deviate from the line, and they tend to be skewed towards brighter and larger fits.}
    \label{fig:unsharpScatter}
\end{figure*}

As discussed in Section \ref{sec:removecompact}, the deep imaging in this work can detect and mask compact objects \textit{on top of} the UDG via unsharp masking. Since shallower studies cannot apply such a mask, we quantify the effect of masking by running \textsc{GALFIT} without this compact objects mask in the \textit{r}-band cutouts. Figure \ref{fig:unsharpScatter} compares the best-fit \textsc{GALFIT} parameters with and without the compact objects mask. Other types of contaminants, bright objects and isolated small objects outside the UDGs, are masked in both cases, as they can be identified in shallower or lower-resolution studies. The presence of compact objects on top of the UDG causes infrequent, but significant, impacts on the \textsc{GALFIT} parameters (e.g., the extreme outliers in Section \ref{subsec:yagiCat}). In a few cases, the presence of compact objects skews the best-fit parameters to much brighter and larger profiles. Among the 1503 UDGs in our catalog, 88 would appear brighter by 0.5 mag and 153 would have $r_{\rm eff}$ larger by 0.5 kpc without the compact object masks, than their counterparts fitted with the masks. 

\subsection{\textit{r}-band Structural parameters}
\label{sec:rBand}

\begin{figure}[htp]
    \centering
    \includegraphics[width=8cm]{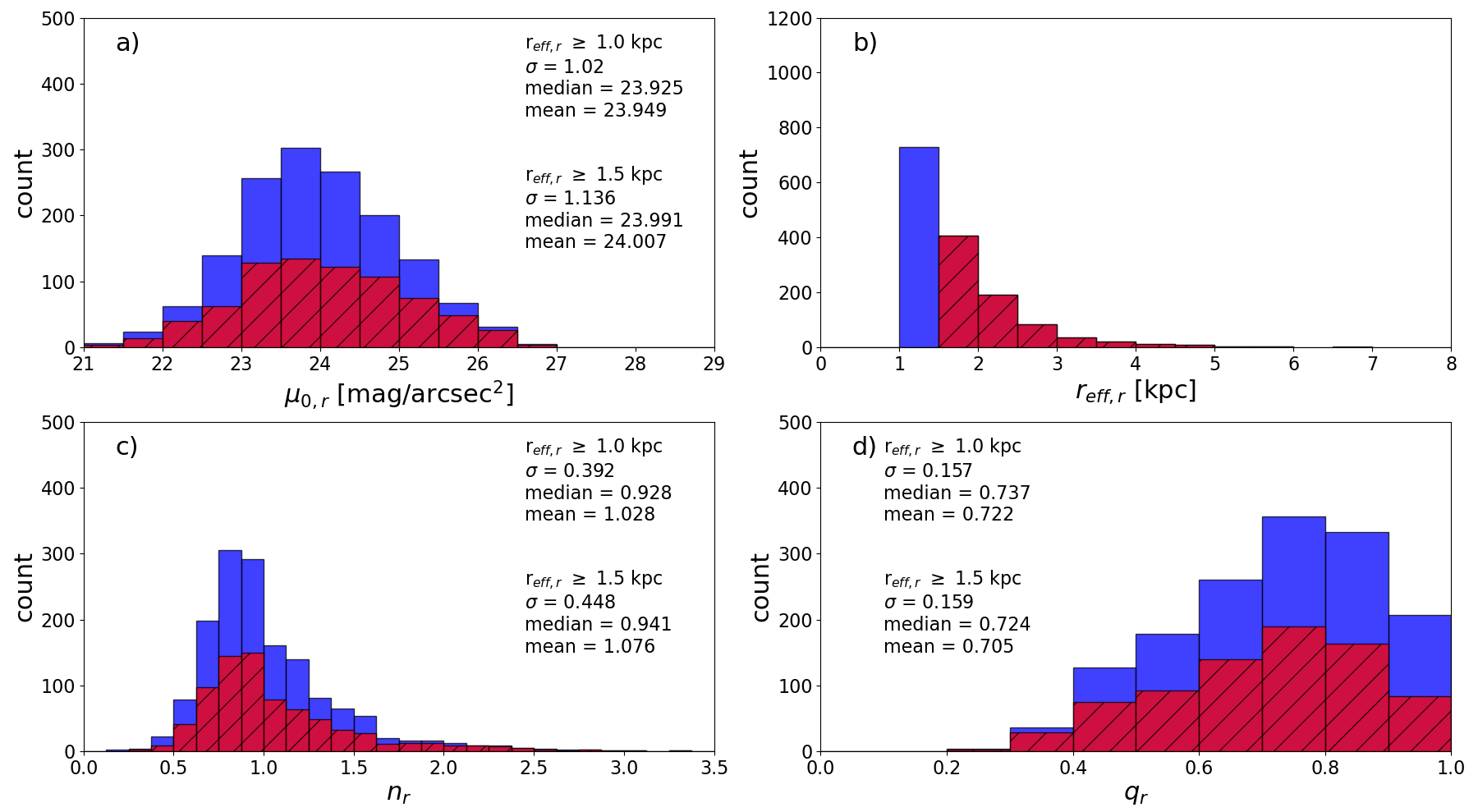}
    \caption{Histograms of \textsc{GALFIT} measured \textit{r}-band (a) $\mu_{\rm 0}$, (b) $r_{\rm eff}$, (c) $n$, and (d) $q$. All UDGs ($r_{\rm eff} \geq 1.0$) are shown in blue, while UDGs with $r_{\rm eff} \geq 1.5$ are overplotted in red.}
    \label{fig:rStructures}
\end{figure}

Figure \ref{fig:rStructures} shows the histograms of \textit{r}-band $\mu_{\rm 0}$, $r_{\rm eff}$, $n$, and $q$ in the \textit{r}-band. We show the results for UDG candidates of all sizes, including sub-UDGs with $1.0 \leq r_{\rm eff, r} < 1.5$, and UDGs with $r_{\rm eff, r} \geq$ 1.5. The average S\'ersic index, $\langle n_{\rm r} \rangle$,for both UDGs and sub-UDGs is between 1.0 and 1.1, similar to dwarf elliptical galaxies, which have comparable stellar mass but are less diffuse. The average axis ratio, $\langle q_{\rm r} \rangle$, is between 0.72 and 0.74 for both UDGs and sub-UDGs. The skew to large $q_{\rm r}$ is not consistent with a population of randomly oriented thin-disk galaxies in a statistical sense, but the \textit{ISOAREA} cut in Section \ref{sec:sourcedetect} may bias our detection against low axis ratios. The relative lack of round ($q_{\rm r} \geq$ 0.9) UDGs compared to the peak favors the interpretation that UDGs are randomly oriented, oblate-triaxial shapes \citep{Rong2020b, KadoFong2021}. 

\begin{figure}[htp]
    \centering
    \includegraphics[width=8cm]{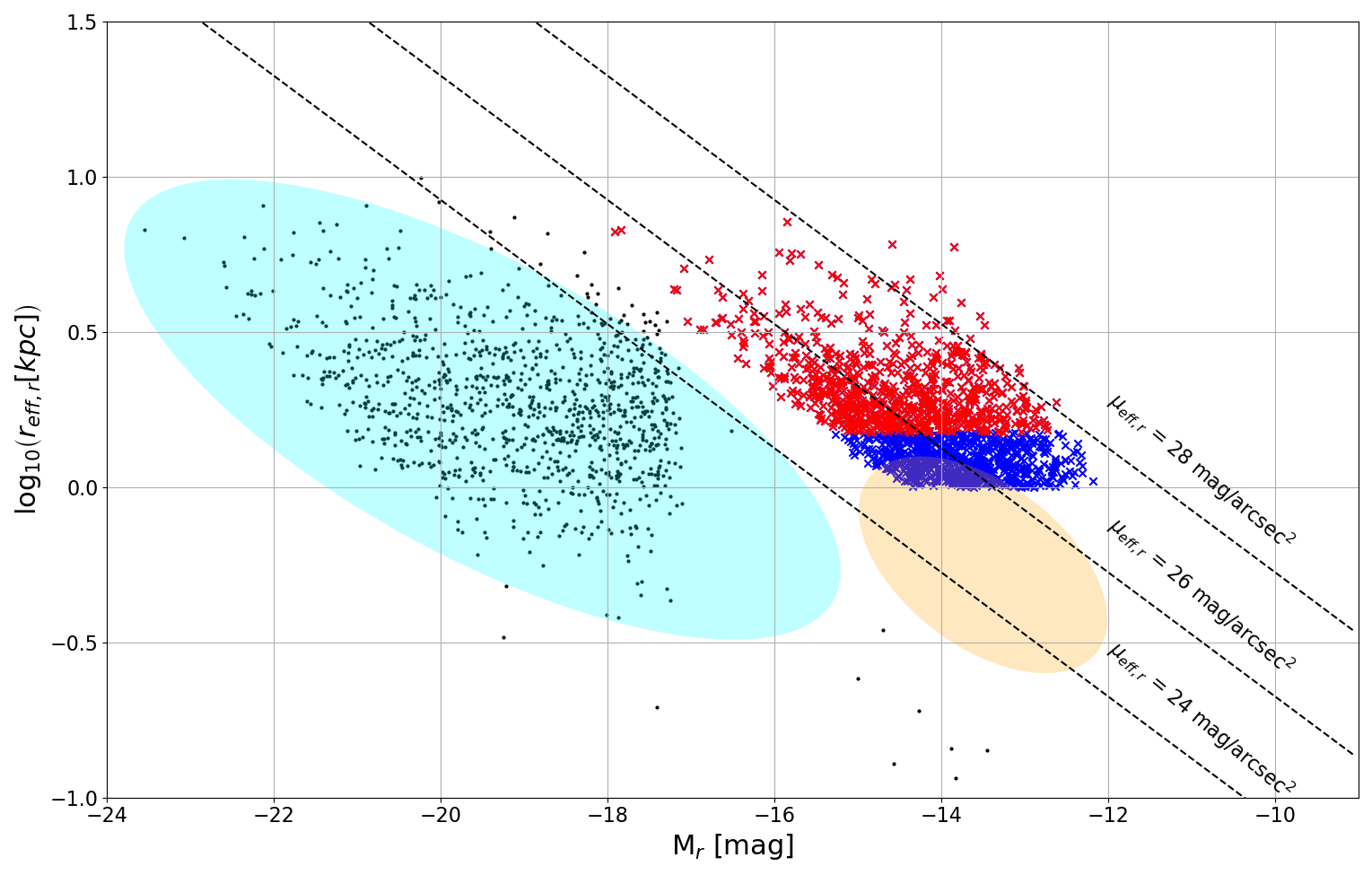}
    \caption{\textsc{GALFIT} $r_{\rm eff, r}$ vs. $m_{\rm r}$ for UDGs using a distance modulus of 35.05 (red crosses for $r_{\rm eff, r} \geq 1.5$, blue crosses for $1.0 \leq$ $r_{\rm eff, r}$ $< 1.5$). For comparison, we plot the $r_{\rm eff, r}$ (\textit{expRad\_R}) vs. $m_{\rm r}$ of SDSS DR17 galaxies (black dots) near the cluster redshift (0.01 $\leq$ z $\leq$ 0.04) in a $3\arcdeg$ radius region around the cluster center are plotted for comparison. The cyan and orange shaded regions show the typical distributions for elliptical and dwarf galaxies, respectively.}
    \label{fig:sbScatter}
\end{figure}

Figure \ref{fig:sbScatter} shows a $r_{\rm e}$-$M_{\rm r}$ plot of the UDGs, with normal galaxies around the Coma cluster from the SDSS DR17 data \citep{SDSSDR17} also plotted as a reference\footnote{The SDSS $m_{\rm r}$ are converted to the equivalent in HSC using equation (\ref{eq5}) in the following section}. Black dashed lines show constant effective surface brightness assuming an exponential profile ($n = 1$). The majority of the UDGs trend towards a region similar to dwarf galaxies (the orange shaded region in Figure \ref{fig:sbScatter}) \citep[$M_{\rm r}$ $\geq$ -15, $r_{\rm eff, r} \leq$ $1.5$, and $n_{\rm r}$ $\approx$ 1.0, ][]{BoselliGavazzi2014}, which suggests UDGs and dwarf galaxies are related, and not distinct populations.

\subsection{g - r Color}
\label{sec:color}

\begin{figure}[htp]
    \centering
    \includegraphics[width=8cm]{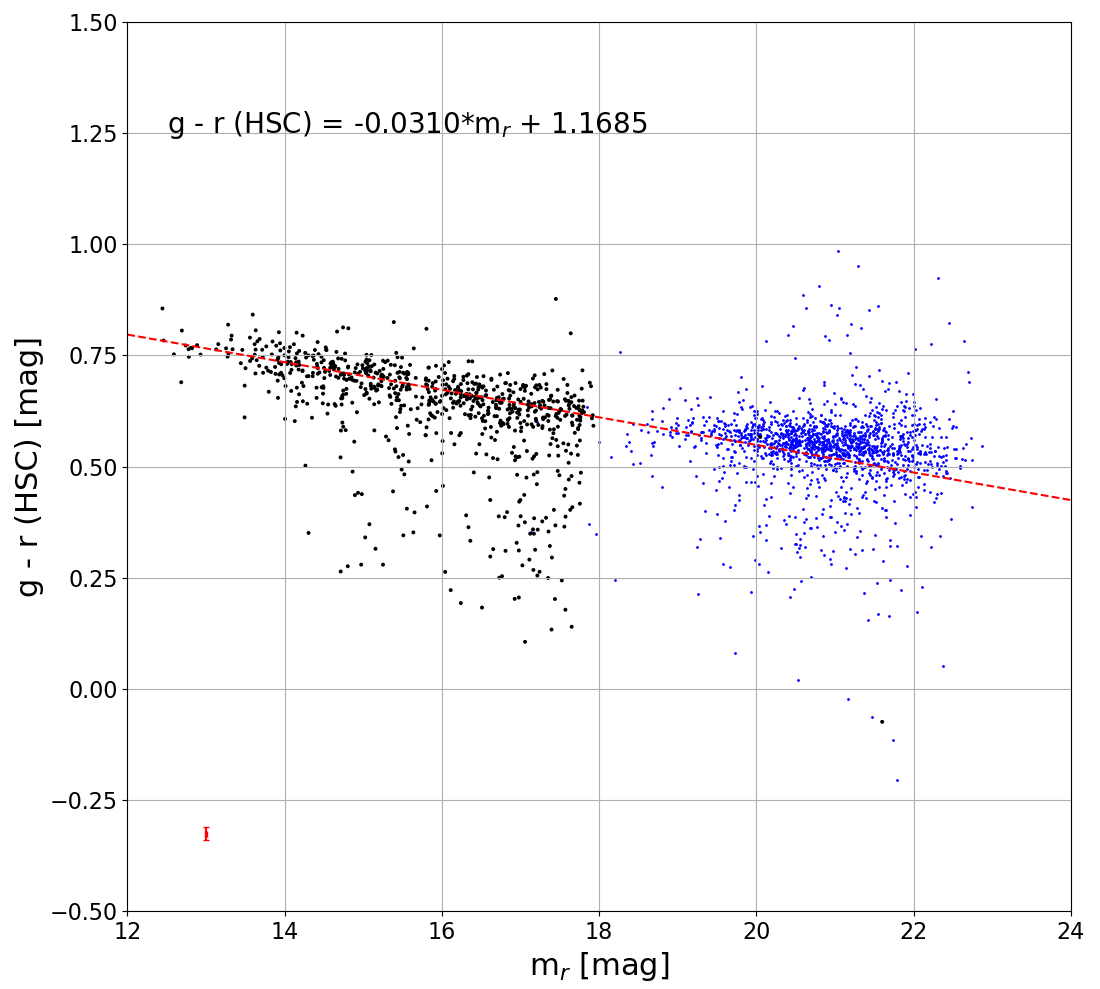}
    \caption{Aperture color-magnitude diagram for UDGs (blue) in this work and SDSS galaxies (black) within a 4$\arcdeg$ $\times$ 4$\arcdeg$ box around the Coma cluster center and in the redshift range $0.018 \leq z \leq 0.028$. The SDSS red sequence (red dashed line) is taken as the linear fit of the peak in the color distribution vs $m_{\rm r}$ per 0.5 mag bin. The red point and error bar shows typical error of this work.}
    \label{fig:cmd}
\end{figure}

\begin{figure}[htp]
    \centering
    \includegraphics[width=8cm]{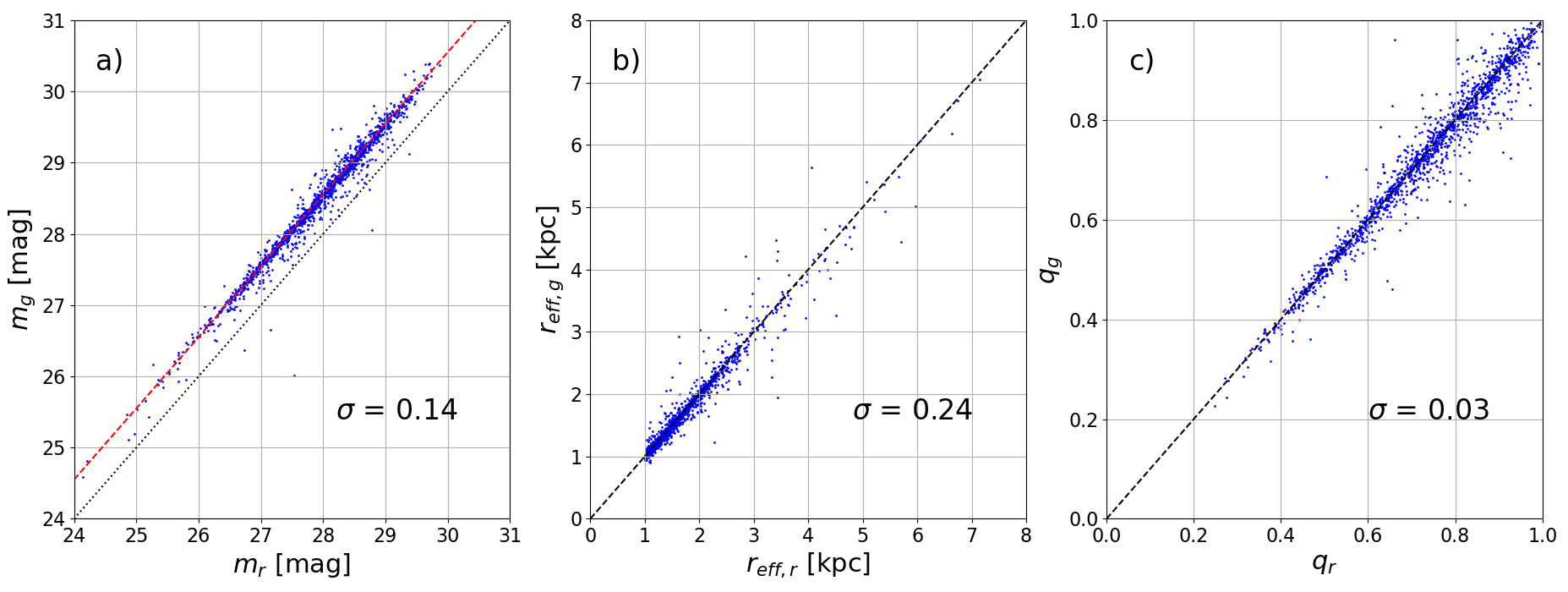}
    \caption{Scatter plot of \textsc{GALFIT} (a) $m$, (b) $r_{\rm eff}$, and (c) $q$ between \textit{g}- and \textit{r}-band values. The S\'ersic index in \textit{g}-band is held fixed to the \textit{r}-band value when performing the fit. Lines of equality (black dashes) are plotted for $r_{\rm eff}$ and $q$, while the red dashed line is shifted by the median color of 0.55.}
    \label{fig:grStructures}
\end{figure}

Figure \ref{fig:grStructures} plots the \textsc{GALFIT} measured structural parameters in \textit{r}-band versus the corresponding value in \textit{g}-band. The structural parameters between the \textit{g}- and \textit{r}-bands are tightly correlated. For fits with equal S\'ersic indices between bands, this suggests that most UDGs have uniform color profiles. We show the aperture color-magnitude diagram of the UDGs in Figure \ref{fig:cmd} along with SDSS galaxies within a redshift range $0.018 \leq z \leq 0.028$ and a $4\arcdeg$ $\times$ $4\arcdeg$ around the Coma cluster center. We also estimate the SDSS red sequence by linearly fitting the peak of the SDSS g - r distribution vs $m_{\rm r}$ per 1 mag bin in the range $13 \leq m_{\rm r} \leq 18$, where the red sequence is well-populated. Using the conversion equations derived with a set of spectra from \citet{Furusawa2000} and transmission curves from \citet{Doi2010}, the SDSS colors are converted to HSC colors as 

\begin{equation} \label{eq5}
    \begin{split}
        g_{\rm HSC} & = g_{\rm SDSS} - 0.0857(g-r)_{\rm SDSS} - 0.0022\\
        r_{\rm HSC} & = r_{\rm SDSS} - 0.0112(g-r)_{\rm SDSS} + 0.0013
    \end{split}
\end{equation}
The UDGs are clustered around the SDSS red sequence extrapolated to fainter magnitudes, indicating they are a quiescent and passively-evolving population.

Figure \ref{fig:cmd} shows very small deviations of the UDGs ($\sim 0.1$ in g-r) from
the red-sequence fit (solid line), although in \citet{Koda2015}, the
UDGs are right on the red-sequence in the B-R vs R plane.
At this point, it is difficult to conclude if the deviations are real.
To confirm such small deviations, we need to analyze the data consistently
for the reference sample that defines the red-sequence.
Currently, the reference sample is from SDSS,
and their parameters are measured in a different scheme and in different bands from ours. 
Note in \citet{Koda2015}, the colors and magnitudes of both their UDGs and reference sample are taken from \citet{Yamanoi2012}, and hence, are based on the same data and analysis. 
In addition, in the range of the UDGs, the solid line is an extrapolation
by almost 4 mag from the fitted range. 
A slight change in its slope could potentially put both the SDSS galaxies
and the UDGs on the same line.

\begin{figure}[htp]
    \centering
    \includegraphics[width=8cm]{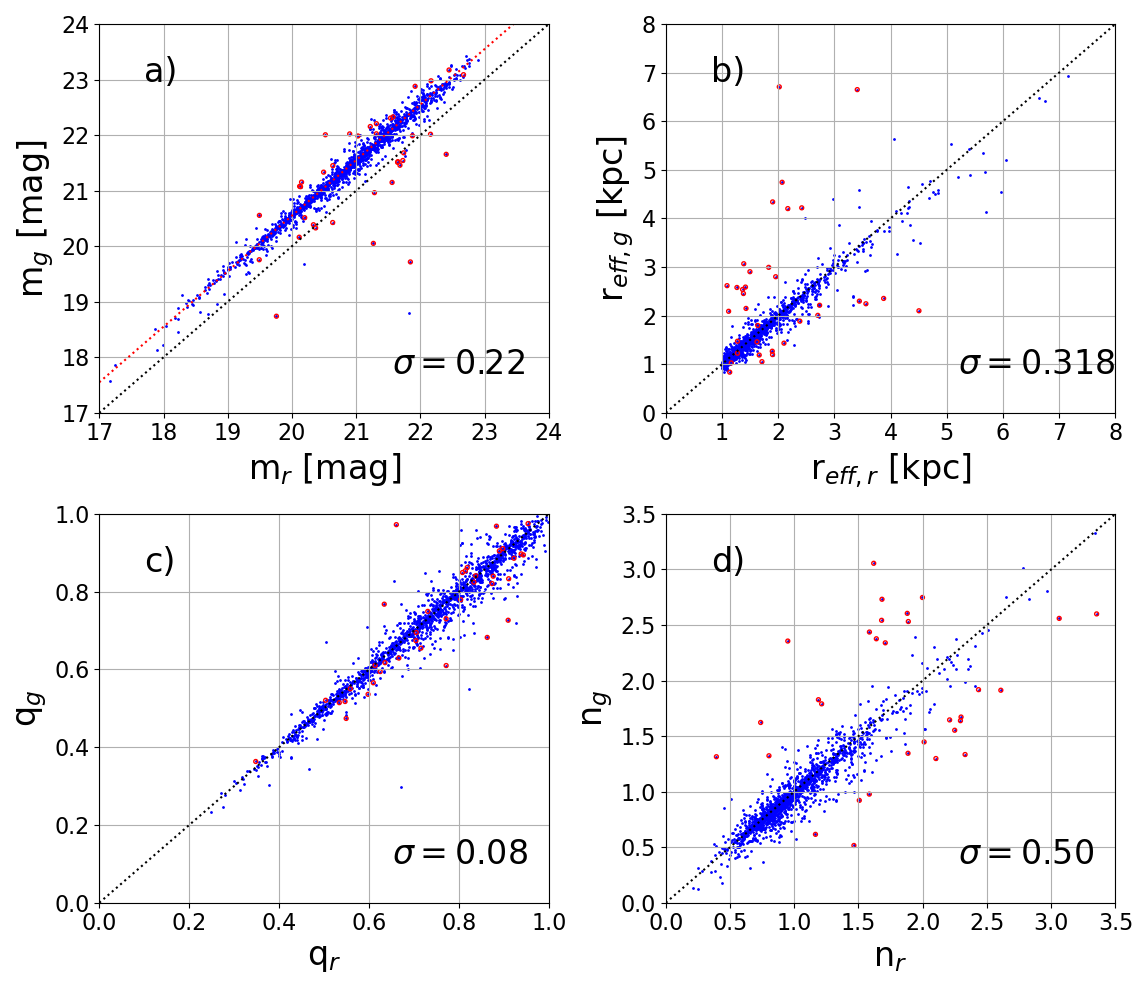}
    \caption{Same as Figure \ref{fig:grStructures}, but with S\'ersic index independent between the two bands. Red dots are 1-$\sigma$ outliers of the S\'ersic index correlation plot. $\sigma$ is calculated relative to the line of unity, except (a) where it is relative to the line shifted by the median color of 0.55.}
    \label{fig:outgrStructures} 
\end{figure}

When measuring the \textit{g}-band parameters, the $n_{\rm g}$ of each UDG was fixed to the corresponding value in \textit{r}-band. Leaving both $n_{\rm g}$ and $n_{\rm r}$ free during the fit instead does not substantially change the correlation, but adds additional scatter (see Figure \ref{fig:outgrStructures}). The S\'ersic indices, despite being fit independently, are correlated between \textit{g}- and \textit{r}-band. Outliers of the S\'ersic index correlation plot also become outliers in the $r_{\rm eff}$ correlation plot relative to the line of unity ($n_{\rm g} = n_{\rm r}$), which may be due to the degeneracy between $r_{\rm eff}$ and S\'ersic index in fitting (Figure \ref{fig:chi2basin}).

\subsection{Nucleated UDGs}

\begin{figure}[htp]
    \centering
    \includegraphics[width=8cm]{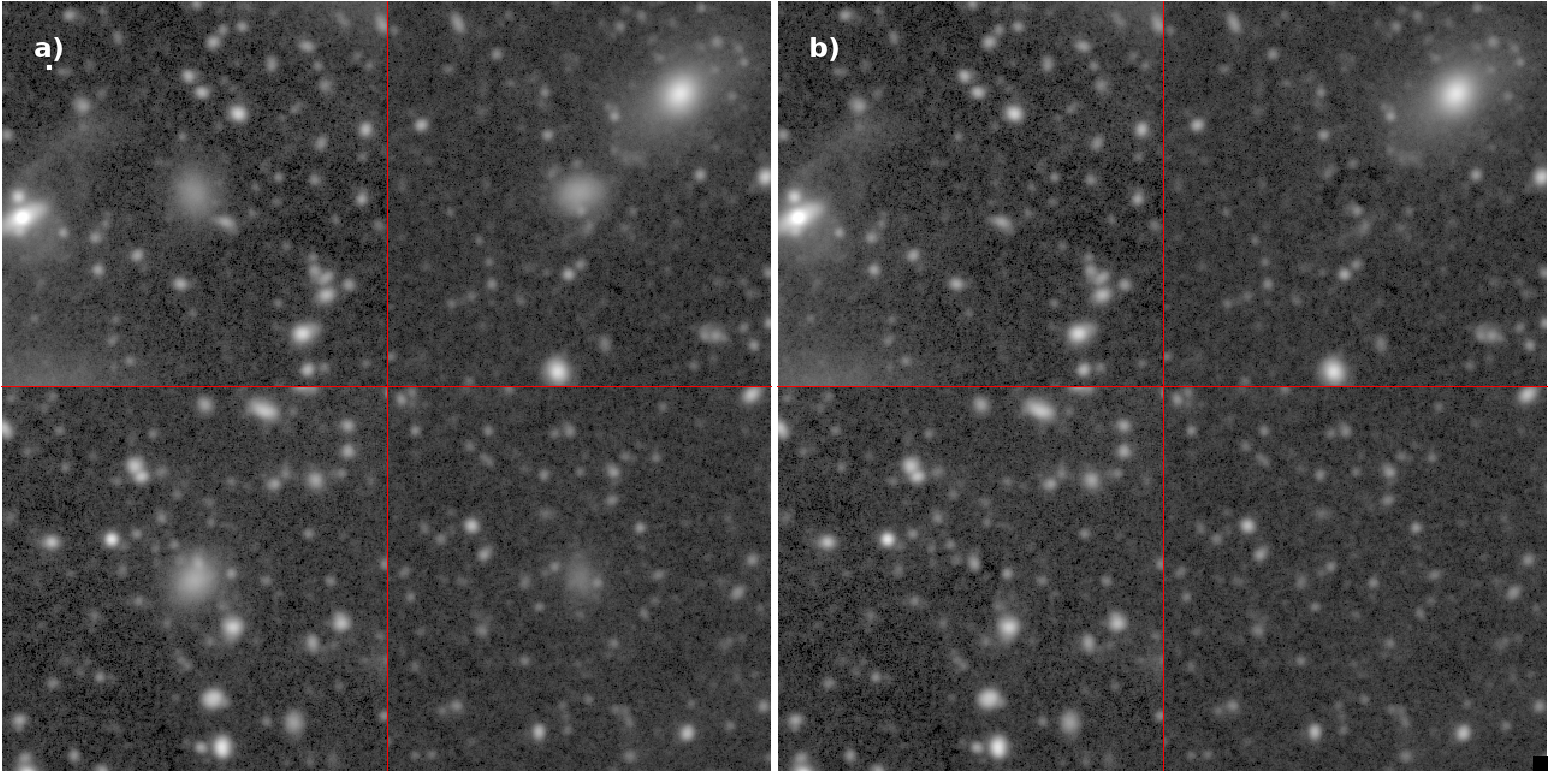}
    \caption{Example UDGs which were classified as nucleated in \citet{Yagi2016} and non-nucleated in this work: a) raw, and b) residual after subtracting model from \textsc{GALFIT}.}
    \label{fig:marginalNuc}
\end{figure}

As mentioned in Section \ref{sec:finalRefinement}, 309 UDGs and sub-UDGs are classified as nucleated, 183 of which are UDGs with $r_{\rm eff} \geq 1.5$ and 126 are sub-UDGs with $r_{\rm eff}=1.0$-$1.5$. 
Our total nucleation fraction ($f_n$), including the sub-UDGs, is therefore 21\%. This fraction is approximately half the value reported of 52\% in \citet{Yagi2016}, which also used BIC to determine which UDGs are nucleated. The main difference between the two catalogs is that our $m_{\rm psf}$ are rarely fainter than 26, whereas \citet{Yagi2016} list $m_{\rm psf}$ as faint as 27.79 for nucleated UDGs in Suprime-Cam. Figure \ref{fig:marginalNuc} shows 4 sample UDGs, which were classified as nucleated in \citet{Yagi2016}, but are re-classified as non-nucleated in this work. If the nucleated UDGs in \citet{Yagi2016} were limited to $m_{\rm psf} \leq 26$ as well, the $f_n$ would be closer to our value, at 21\%. 

\begin{figure}[htp]
    \centering
    \includegraphics[width=8cm]{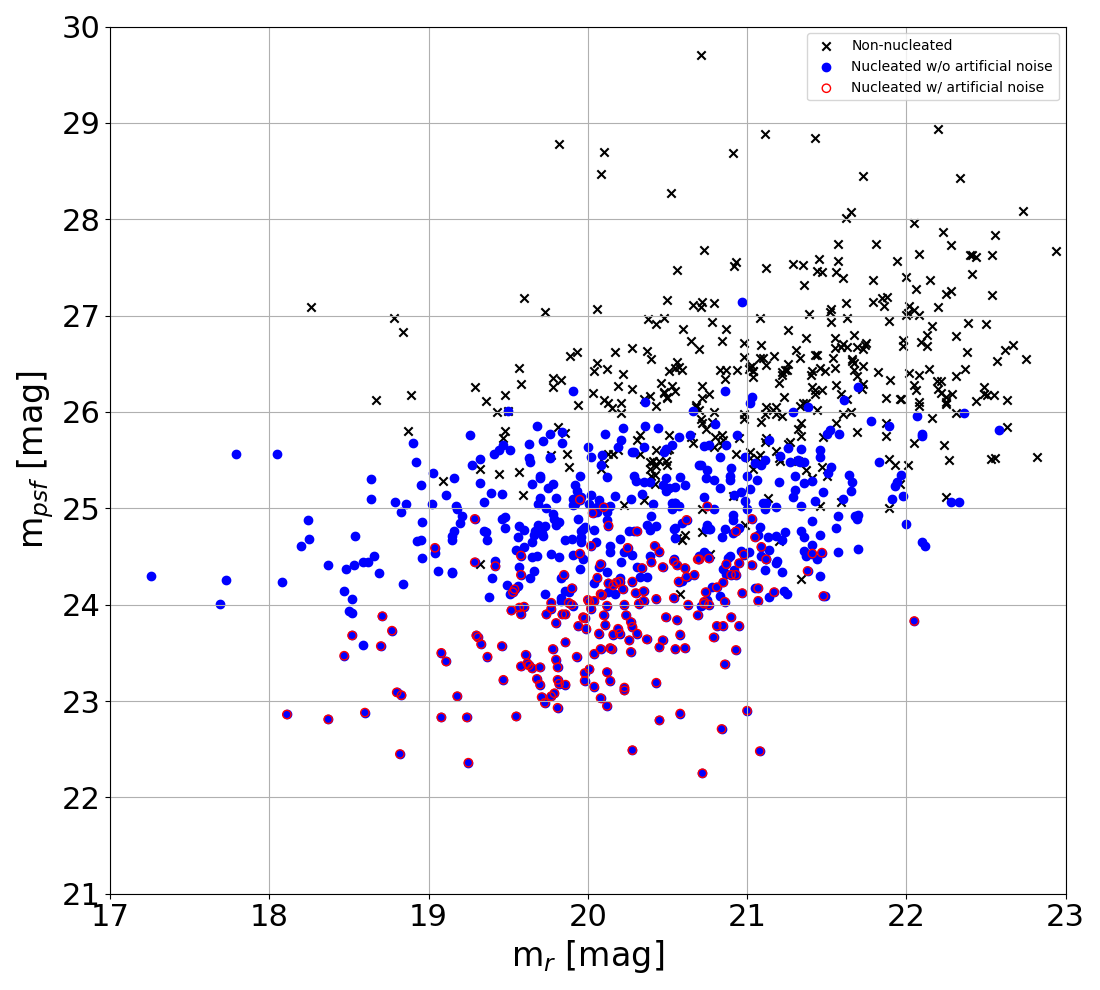}
    \caption{Magnitudes of \textit{r}-band PSF components vs S\'ersic components from S\'ersic + PSF fitting. UDGs judged as non-nucleated by BIC are shown as black crosses, and nucleated UDGs are shown as blue dots. The $m_{\rm psf}$ of the latter are almost completely confined to $m_{\rm psf} \lesssim 26.0$. When additional noise is included in the image before fitting, UDGs that are still classified as nucleated are marked with red open circles. With additional noise in the image, the upper limit on $m_{\rm psf}$ has dropped to $m_{\rm psf} \lesssim 25.0$.}
    \label{fig:psfMagnitudes}
\end{figure}

The effective cutoff in $m_{\rm psf}$ is tied to the BIC classification and the estimated noise level of the image. To illustrate this, we increased the noise in each cutout by adding artificial noise equal to the on-sky noise and re-fit the UDG before redetermining whether they are nucleated or not. In Figure \ref{fig:psfMagnitudes}, we plot the magnitude of the PSF component vs the magnitude of the S\'ersic component for all convergent S\'ersic + PSF fit results, separated by whether the UDG is nucleated or not. When artificial noise is added, the upper limit on $m_{\rm psf}$ decreases from 26 to 25. This shows that the nucleated/non-nucleated classification is affected by signal-to-noise. In \citet{Yagi2016}, an incorrectly low background noise was used, which resulted in picking up faint and insignificant PSF components. The differences in the sky-subtraction and masking, discussed in Section \ref{subsec:yagiCat}, also contribute to the classification, but their effects are relatively minor and reduce the nucleation fraction by about 5\%.

\begin{figure}[htp]
    \centering
    \includegraphics[width=8cm]{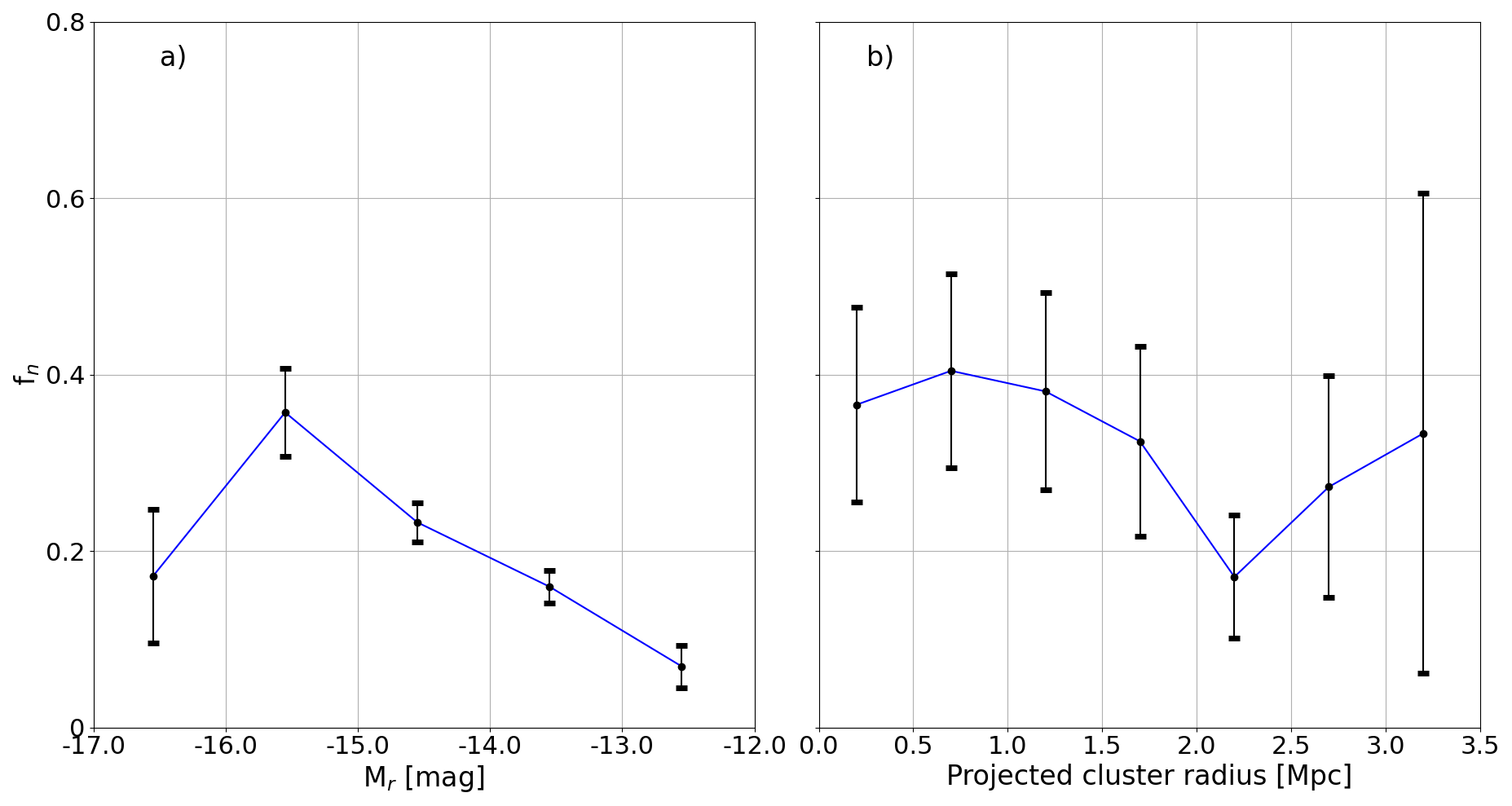}
    \caption{Nucleation fraction $f_{\rm n}$ as a function of a) $M_{\rm r}$ and b) projected radius with respect to the Coma cluster center. The latter is restricted to $m_{\rm r} \leq 20$ ($M_{\rm r} \leq -15.05$) to avoid incompleteness from the limit in $m_{\rm psf}$. The error bars assume Poissonian noise.}
    \label{fig:nucleationMag}
\end{figure}

Figure \ref{fig:nucleationMag}a shows the nucleation fraction as a function of absolute magnitude $M_{\rm r}$ for our UDGs. We note that in the literature, the brightest and most massive dwarf galaxies can reach $f_n$ close to 1, decreasing with fainter luminosities \citep{denBrok2014, Munoz2015, Hoyer2021}. In our analysis, the decrease in $f_n$ with fainter $M_{\rm r}$ is the result of the bias with the method of nucleation classification we have chosen. We use the BIC to determine whether a UDG is nucleated or not, and this appears to impose a limit on $m_{\rm psf}$ that depends on the signal-to-noise ratio. As our $m_{\rm psf}$ are limited to values brighter than 26, and $m_{\rm r}$ is mildly correlated with $m_{\rm psf}$ in Figure \ref{fig:psfMagnitudes}, $f_n$ apparently decreases at fainter $M_{\rm r}$. However, it is also possible that nucleation fraction may be correlated with luminosity generally, due to more massive galaxies having a deeper potential and thus a better ability to pile material in their central regions \citep{Sanchez-Janssen2019, Zanatta2021}. 

In Figure \ref{fig:nucleationMag}b, we show $f_n$ as a function of projected cluster radius for $m_{\rm r} \leq 20$. As discussed above, our $f_{\rm n}$ does not include nuclei fainter than $m_{\rm psf} = 26.0$, so we exclude UDGs fainter than 20 to avoid incompleteness from this limit in $m_{\rm psf}$ based on the behavior in Figure \ref{fig:psfMagnitudes}. We find the nucleation fraction of Coma UDGs is approximately constant across the entire cluster. This is in contrast with the literature, where quiescent nucleated galaxies that contain nuclear star clusters appear to be preferentially located in denser environments, such as the centers of clusters \citep{Baldassare2014, Lim2018, Ordenes-Briceno2018, Sanchez-Janssen2019, Zanatta2021}. The lack of radial trend in nucleated UDGs, as opposed to clear trends in other galaxy populations, may have an implication on their formation history.

\subsection{Spatial Distribution}

\begin{figure}[htp]
    \centering
    \includegraphics[width=8cm]{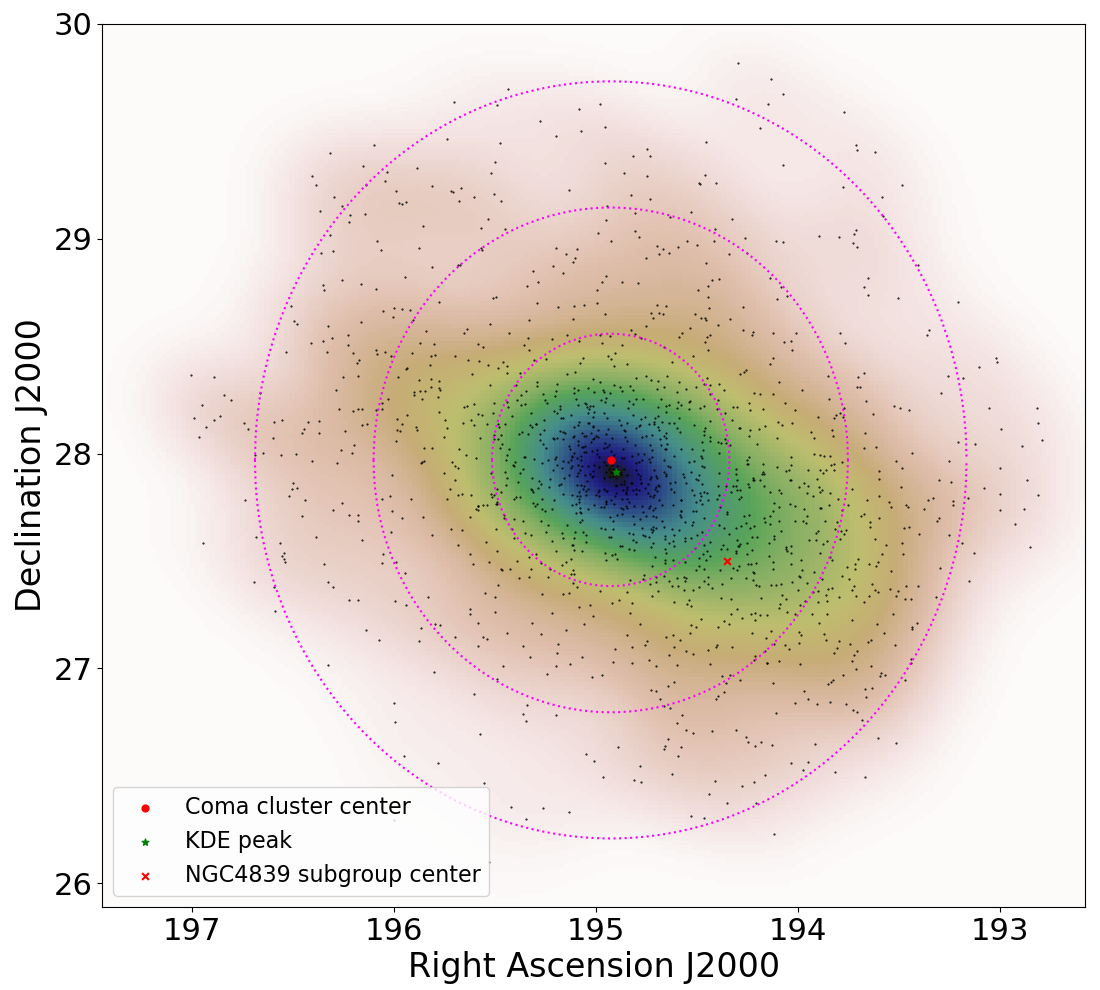}
    \caption{Kernel density estimation of UDG position, with a Gaussian kernel. Dashed magenta lines mark increments of 1 Mpc in projected cluster radius}
    \label{fig:kde}
\end{figure}

\begin{figure}[htp]
    \centering
    \includegraphics[width=8cm]{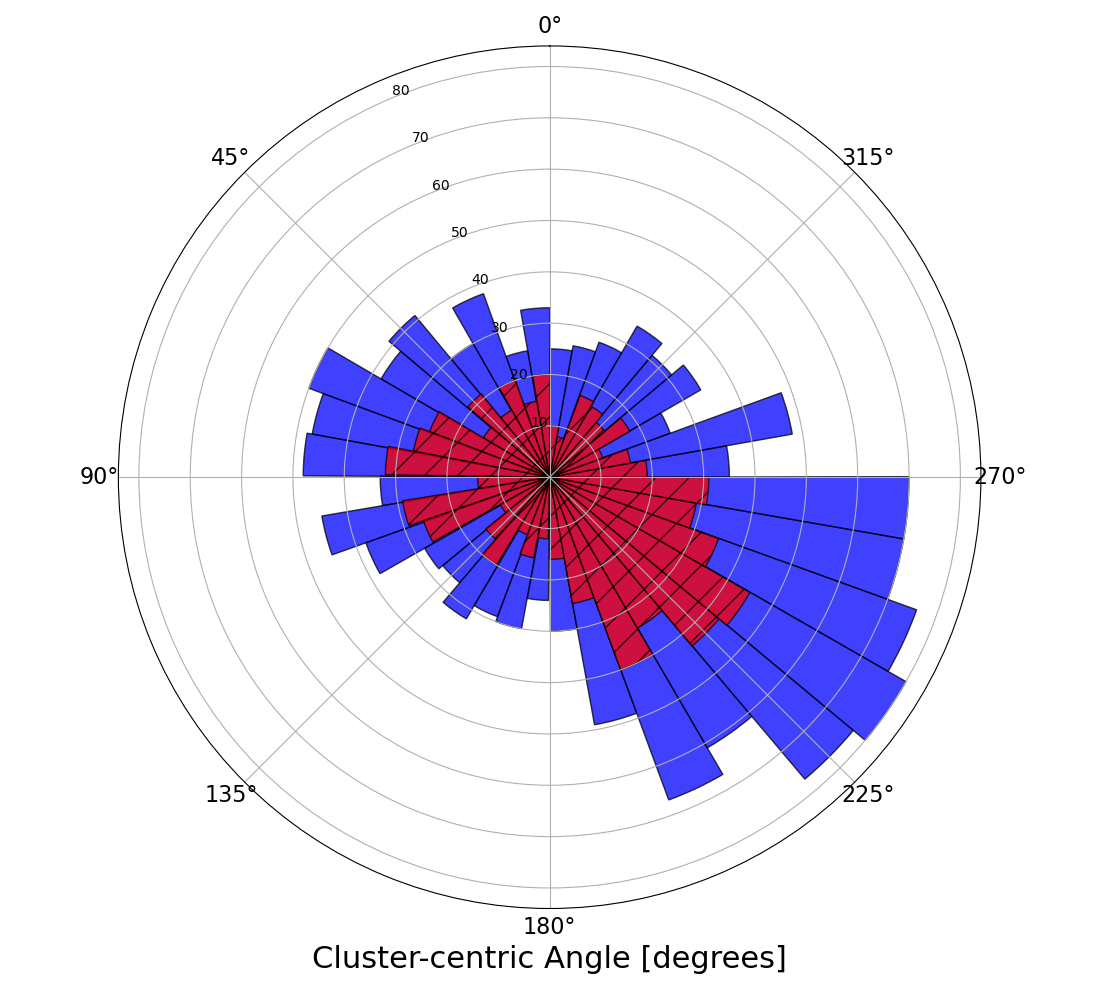}
    \caption{Angular histogram of all UDGs in the catalog (blue) and UDGs with $r_{\rm eff, r} \geq 1.5$ (red) in 10$\arcdeg$ bins.}
    \label{fig:angularDist}
\end{figure}

Figure \ref{fig:kde} is another presentation of the UDG distribution (from Figure \ref{fig:radecScatter}). We show the kernel density estimator of the UDGs positions using a Gaussian kernel in Figure \ref{fig:kde}. The kernel bandwidth is $0.\arcdeg297$, estimated through Scott's rule. There is a clear elongation to the distribution. To better quantify the anisotropy, we take a circular histogram by binning along cluster-centric angle by a $10\arcdeg$ increment (see Figure \ref{fig:angularDist}) centered at the Coma cluster center $(\alpha_{\rm J2000}, \delta_{\rm J2000}) =$ (12:59:42.8, +27:58:14). There are 2 peaks in two directions ($230\arcdeg$ and $70\arcdeg$), which coincide with the direction of the Coma filaments \citep{Fontanelli1984} connecting the cluster to the large scale structure. The southwestern ($230\arcdeg$) peak is in the direction of the NGC4839 subgroup and may connect to the southwestern filament \citep{Akamatsu2013} going to A1367, while the northeastern peak ($70\arcdeg$) is in the direction of a filament going to A2199. \citet{Malavasi2020} report a potential third filament towards the north, but there is no noticeable corresponding excess in our sample of UDGs.

The alignment of the UDGs distribution along the large scale structure around Coma reassures the interpretation that most of them, from the center to the outskirts, lie at the distance of the Coma cluster. Conversely, this means UDGs should trace out the large scale structure, and be fairly common in filaments \citep{RomanTrujillo2017, RomanTrujillo2017b}. 

\begin{figure}[htp]
    \centering
    \includegraphics[width=8cm]{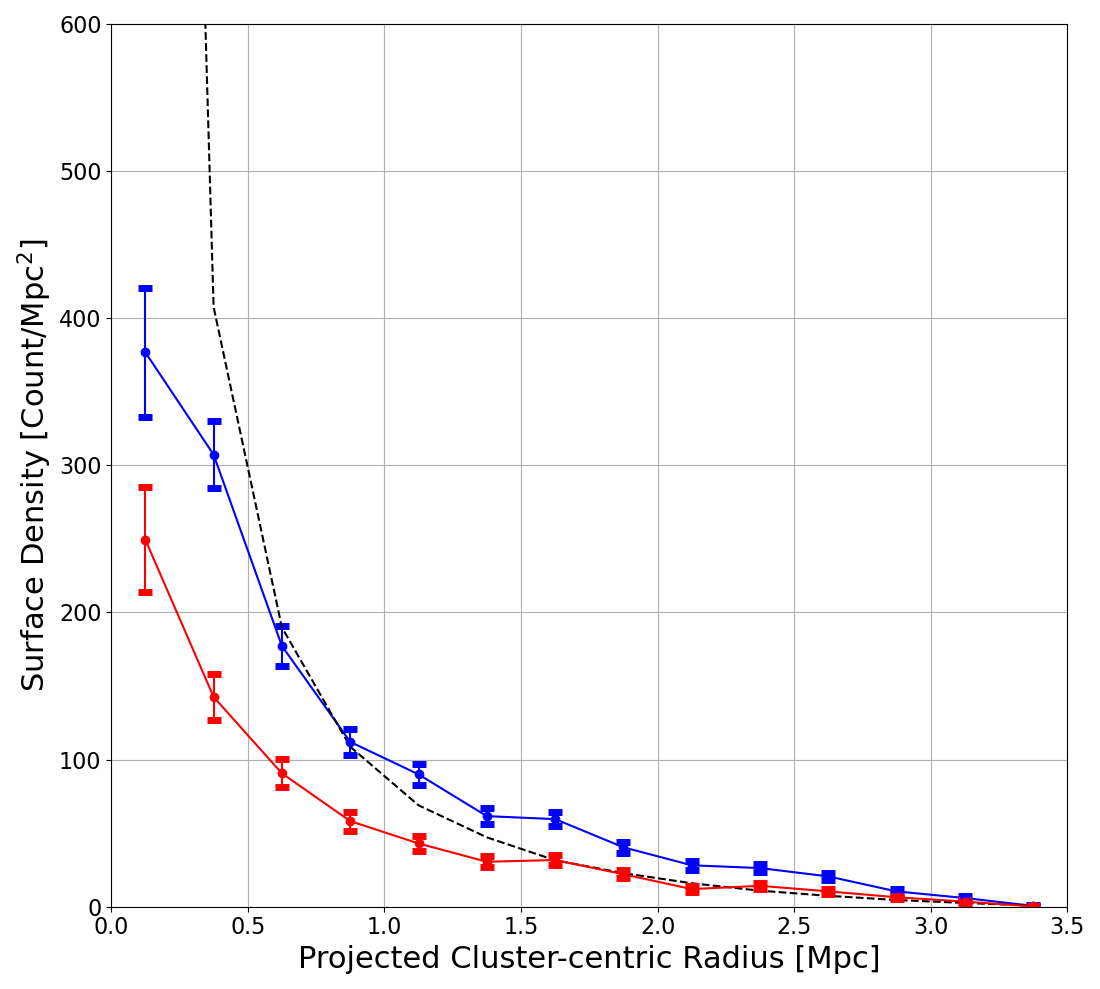}
    \caption{Surface density of all UDG candidates in the catalog (blue) and UDG candidates excluding sub-UDGs (red) in 0.25 Mpc projected radius bins. As a reference, the black dashed line shows the theoretical density if the UDGs were distributed as a uniform sphere.}
    \label{fig:radialDist}
\end{figure}

Figure \ref{fig:radialDist} shows the projected radial surface density of all 1503 UDGs, and the subset of 774 UDGs with $r_{\rm eff, r} \geq 1.5$. The two profiles are similar, with the former being approximately double the latter at all projected radii. In combination with Figure \ref{fig:angularDist}, we see that sub-UDGs ($r_{\rm eff}=1$-$1.5$) share the same spatial distribution as the UDGs ($r_{\rm eff} \geq 1.5$). At approximately 0.5 Mpc there is an inflection point in the projected radial surface density plot. If the UDGs followed the distribution of a uniform sphere, the projected surface density would be concave up everywhere, so there are less detected UDGs in the very central region than a spherical distribution predicts. UDG may be destroyed by tidal forces below a certain radius \citep{PVD2015}. However, the lack of central UDGs could also be due to either the diffuse light in the cluster center affecting the background subtraction and making the detection more difficult, or obscuration by bright galaxies \citep{Adami2006}.

\section{Conclusion}
\label{sec:Conclusion}
We present an updated catalog of UDGs in the Coma cluster in \textit{g}- and \textit{r}-band. We have outlined an automated procedure to detect and measure UDGs in the presence of contaminants. This procedure uses multiple \textsc{SExtractor} runs to remove specific types of contaminants in images before searching for UDGs. Our main results are the following:

\begin{itemize}

    \item We develop a cleaning algorithm to resolve the problem of foreground and background object interfering with the detection of UDGs. We search for UDGs across the whole Coma cluster out to the virial radius, approximately 3 Mpc from the cluster center, and the surrounding area beyond the virial radius. Our larger coverage has increased the number of UDGs to 1503. Among them, 774 have $r_{\rm eff, r} \geq 1.5$. We also measure the colors of all the UDGs.
    
    \item The new UDGs show internal properties consistent with those of the previous studies (e.g., S\'ersic index of $\sim$1), and are distributed across the cluster, with a concentration around the cluster center. With the addition of \textit{g}-\textit{r} color, we see that the $r_{\rm eff}$ and $\langle \mu \rangle_{\rm eff}$ between \textit{g}- and \textit{r}-bands are tightly correlated, while their colors place them around the red sequence of Coma. They are seen to be a passively-evolving population, around the red sequence of Coma. 
    
    \item The whole cluster coverage reveals that the spatial distribution of UDGs aligns with the large-scale structure. This supports the interpretation that most of our UDGs are cluster members. The excess toward the south-west direction from the cluster center also coincides with the location of the NGC4839 subgroup, so many UDGs in this direction are likely to be associated with this infalling group.
    
\end{itemize}

\newpage
This work was built upon and expanded our previous study \citep{Yagi2016}. All, or parts, of our method could be combined with other potential approaches, e.g., \textit{NoiseChisel} \citep{Akhlaghi2015} and others, for potentially better detection of extended LSB objects. While testing the other suggested approaches is beyond the scope of this work, we expect further improvement on the detection and measurement method, and hence on studies of extended LSB objects, in particular in view of the upcoming Vera Rubin survey.

\acknowledgments
% Individuals
We appreciate the editor and anonymous referee for thorough reading and constructive comments.
We thank Samuel Boissier and Junais for helpful discussions.
We thank Fumiaki Nakata, the support astronomer at Subaru, and the HSC helpdesk for helping our observations. 
% NAOJ & ADC
This research is based on data collected at the Subaru Telescope, which is operated by the National Astronomical Observatory of Japan (NAOJ). We are honored and grateful for the opportunity of observing the Universe from Maunakea, which has the cultural, historical, and natural significance in Hawaii.
Data analysis is in part carried out on the Multi-wavelength Data Analysis System operated by the Astronomy Data Center (ADC) in NAOJ.
% NED
\footnote{https://ned.ipac.caltech.edu/}.
This research has made use of the NASA/IPAC Extragalactic Database (NED), which is operated by the Jet Propulsion Laboratory, California Institute of Technology, under contract with the National Aeronautics and Space Administration.
% SDSS
This work has made use of the Sloan Digital Sky Survey (SDSS) DR17 archive \footnote{https://www.sdss.org/dr17/}.
Funding for the SDSS IV has been provided by the Alfred P. Sloan Foundation, the U.S. Department of Energy Office of Science, and the Participating Institutions. 
SDSS-IV acknowledges support and resources from the Center for High Performance Computing  at the University of Utah. The SDSS website is www.sdss.org.
SDSS-IV is managed by the Astrophysical Research Consortium for the Participating Institutions of the SDSS Collaboration including 
the Brazilian Participation Group, the Carnegie Institution for Science, Carnegie Mellon University, Center for Astrophysics | Harvard \& Smithsonian, the Chilean Participation Group, the French Participation Group, Instituto de Astrof\'isica de 
Canarias, The Johns Hopkins University, Kavli Institute for the Physics and Mathematics of the 
Universe (IPMU) / University of Tokyo, the Korean Participation Group, Lawrence Berkeley National Laboratory, Leibniz Institut f\"ur Astrophysik Potsdam (AIP),  Max-Planck-Institut f\"ur Astronomie (MPIA Heidelberg), Max-Planck-Institut f\"ur 
Astrophysik (MPA Garching), Max-Planck-Institut f\"ur Extraterrestrische Physik (MPE), National Astronomical Observatories of China, New Mexico State University, New York University, University of Notre Dame, Observat\'ario Nacional / MCTI, The Ohio State University, Pennsylvania State University, Shanghai Astronomical Observatory, United Kingdom Participation Group, Universidad Nacional Aut\'onoma de M\'exico, University of Arizona, University of Colorado Boulder, University of Oxford, University of Portsmouth, University of Utah, University of Virginia, University of Washington, University of Wisconsin, Vanderbilt University, and Yale University.
% Funding
We acknowledge support from NSF through grants AST-1812847 and AST-2006600.

\facilities{Subaru, SDSS, IRSA}
\software{hscPipe 4.0.1 \citep{Bosch2018}, Astropy \citep{astropy:2013, astropy:2018}, dustmaps \citep{Green2018}, \textsc{GALFIT} 3.0.5 \citep{Peng2002, Peng2010}, \textsc{SExtractor} 2.19.5 \citep{BertinArnouts1996}, \textsc{PSFEX} 3.22.1 \citep{Bertin2013}}

%\bibliography{library.bib}
\bibliography{main.bbl}
\newpage

\begin{figure}
    \centering
    \includegraphics[angle=90,origin=c, width = 11cm]{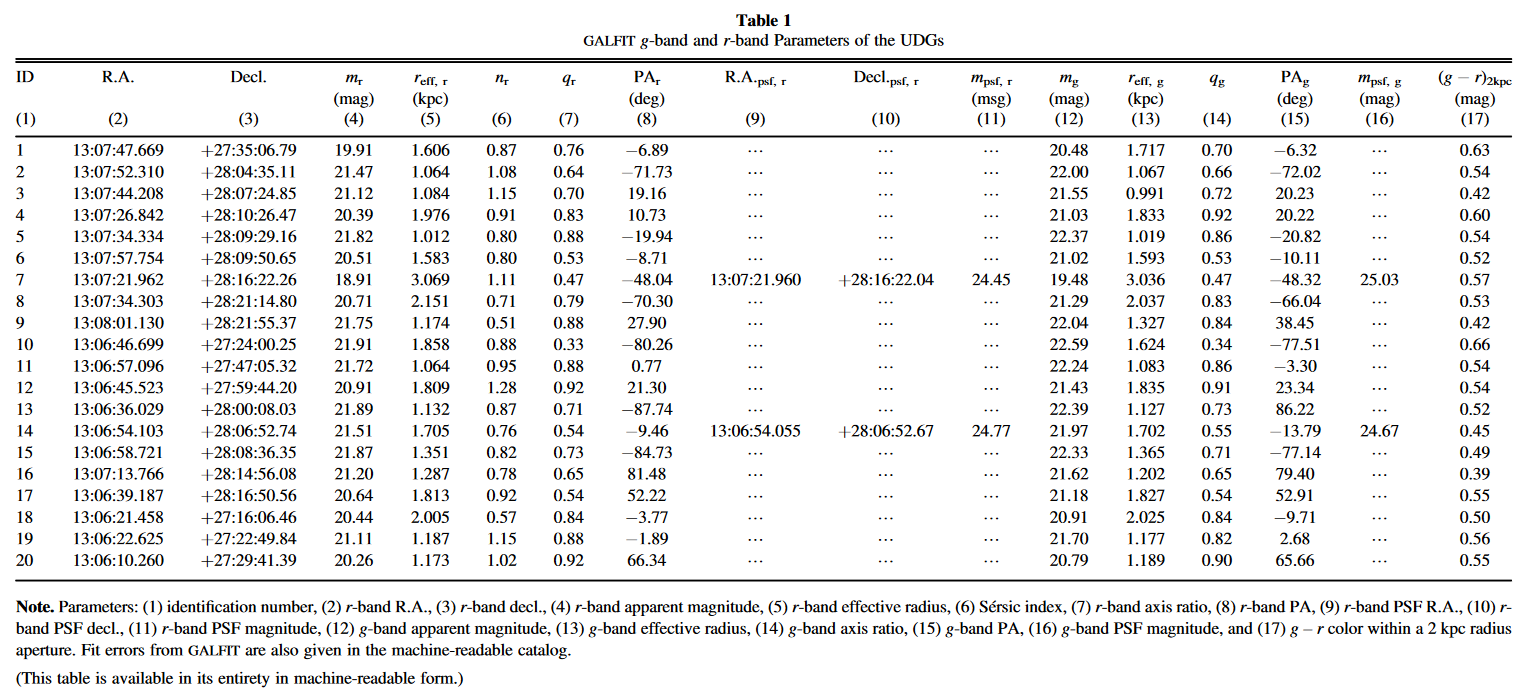}
    \label{table1}
\end{figure}

\end{document}